\documentclass[aps,prb,twocolumn,amsmath,amssymb,superscriptaddress,scrartcl,eqsecnum,longbibliography,nofootinbib]{revtex4-2}

\usepackage[T1]{fontenc}
\usepackage[latin9]{inputenc}
\usepackage{color}
\usepackage{verbatim}
\usepackage{amsmath}
\usepackage{amssymb}
\usepackage{graphicx}

\usepackage{bm}

\usepackage{tikz}
\usepackage{diagbox}
\usepackage[normalem]{ulem}

\makeatletter
\PassOptionsToPackage{caption=false}{subfig} 
\usepackage{hyperref}
\hypersetup{
breaklinks=true,
colorlinks=true,
citecolor=blue,
linkcolor=black,
filecolor=black,
urlcolor=blue
}
\IfFileExists{lmodern.sty}{\usepackage{lmodern}}{}

\renewcommand{\tilde}{\widetilde}
\renewcommand{\hat}{\widehat}
\renewcommand{\leq}{\leqslant}

\renewcommand{\Re}{\operatorname{Re}}
\renewcommand{\Im}{\operatorname{Im}}

\newcommand{\Tr}{\operatorname{Tr}}

\definecolor{darkred}{rgb}{0.5,0.,0.}

\makeatletter

\newcommand*{\wideboxed}[1]{\setlength{\fboxsep}{1ex}%
  \fbox{\m@th$\displaystyle#1$}}
\makeatother

\def\be{\begin{equation}}
\def\ee{\end{equation}}

\makeatother

\begin{document}


\title{Dissipation meets conformal interface in open quantum systems: How the relaxation rate is suppressed}

\author{Ruhanshi Barad}

\affiliation{School of Physics, Georgia Institute of Technology, Atlanta, GA 30332, USA}

\author{Qicheng Tang}

\affiliation{School of Physics, Georgia Institute of Technology, Atlanta, GA 30332, USA}

\author{Xueda Wen}

\affiliation{School of Physics, Georgia Institute of Technology, Atlanta, GA 30332, USA}

\begin{abstract}
Conformal interfaces play an important role in quantum critical systems. 
In closed systems, the transmission properties of conformal interfaces are typically characterized by two quantities: One is the effective central charge $c_{\text{eff}}$, which measures the amount of quantum entanglement through the interface, and the other is the transmission coefficient $c_{\text{LR}}$, which measures the energy transmission through the interface. 
In the present work, to characterize the transmission property of conformal interfaces in \textit{open} quantum systems, we propose a third quantity $c_{\text{relax}}$, which is defined through the ratio of Liouvillian gaps with and without an interface. Physically, $c_{\text{relax}}$ measures the suppression of the relaxation rate towards a steady state when the system is subject to a local dissipation. 
We perform both analytical perturbation calculations and exact numerical calculations based on a free fermion chain at the critical point. 
It is found that $c_{\text{relax}}$ decreases monotonically with the strength of the interface. In particular, 
$0\le c_{\text{relax}}\le c_{\text{LR}}\le c_{\text{eff}}$, where the equalities hold if and only if the interface is totally reflective or totally transmissive.  
Our result for $c_{\text{relax}}$ is universal in the sense that $c_{\text{relax}}$ is independent of (i) the dissipation strength in the weak dissipation regime and (ii) the location where the local dissipation is introduced. When local dissipation is introduced at the boundary, we analytically show that the result holds for any finite dissipation strength; when dissipation is introduced in the bulk or near the interface, we perform a perturbative analysis to establish the relation's validity in generic settings. Comparing to the previously known $c_{\text{LR}}$ and $c_{\text{eff}}$ in a closed system, our $c_{\text{relax}}$ shows a distinct behavior as a function of the interface strength, suggesting its novelty to characterize conformal interfaces in open systems and offering insights into critical systems under dissipation.


\end{abstract}
\maketitle


\section{Introduction}

Impurities are commonly present in real physical systems and have a significant impact on both their equilibrium and non-equilibrium properties ~\cite{Anderson1958localization, anderson1979, kondo1964, mirlin2007review, basko2005, huse2007, huse2014review, Serbyn2018review}. In critical systems, when the impurity itself is a scale-invariant object (a conformal defect/interface), it not only encodes the universal properties of the bulk criticality, but also gives rise to fascinating phenomena that are invisible in the bulk~\cite{Affleck_1992_BCFT_Kondo, Affleck1994folding, Affleck1995_bcft_kondo, oshikawa1997_Ising_defect, Bachas_2002_permeable, Quella2002_sb, Frohlich2004_KW, Quella2006_transmission, meineri2016}. Such impurities have attracted considerable attention due to their conceptual contribution to various fields, including the study of the Kondo effect~\cite{kondo1964, Affleck_1992_BCFT_Kondo, Affleck1995_bcft_kondo}, the development of renormalization group flows~\cite{Brunner_2008_RG_DW, Gaiotto2012_RG_DW, Konechny_2014_RG_defect, Brunner_2016_transmission, Cardy2017_boundary_bulkRG, Konechny2021_RGinterface, Cuomo2021_RG_DefectLine}, the understanding of black hole evaporation~\cite{Myers2020_defect_island_1, Myers2020_defect_island_2, Sonner2022_Island}, and applications to measurement problems~\cite{Rajabpour2015_measurement_cft, Watanabe2016_project_holography, Swingle2022_holo_measure, Popov2022_MIPT_holo, Swingle2022_holo_measure_2, Swingle2023_holo_measure, Altman2022_measure_critical, Weinstein2023, JianCM2023_measure_ising, Alicea2023_measure_ising, JianSK_2023_holo_weak_measure, JianSK2023_measure_LL, Wei2023_MIPT_holo, Alicea2024_teleport_critical, oshikawa2024, Tang2024, LiuYue2024, Zhang2025}, among others.

Because a conformal interface introduces no intrinsic length scale, the reflection and transmission coefficients are independent of the magnitude of the incident momentum or energy. This independence is a useful microscopic diagnostic of scale invariance of a conformal interface and, in closed quantum critical systems, provides a characterization via the transmission coefficient $c_{\text{LR}}$~\cite{Quella2006_transmission, meineri2016, Brunner_2016_transmission, 2020_Meineri, Bachas2020, Bachas2022}, which quantifies energy transmitted across the interface. Another widely used characterization of conformal interfaces in closed system (mainly 1+1-dimensional) is the effective central charge $c_{\text{eff}}$ extracted from entanglement entropy, which measures the amount of information transmitted across the interface~\cite{Kazuhiro_Sakai_2008, Eisler2010, Eisler_2012, Calabrese_2012_EE_junction, brehm_2015_EE_interface_ising, Wen2017, Eisler2022, Tang2023, Karch2023_effective, Karch2024}.
Both $c_{\text{eff}}$ and $c_{\text{LR}}$ are universal in characterizing the interface. For a general 1+1 dimensional conformal field theories (CFTs), no matter the system is in equilibrium or out of equilibrium, it is found that the quantum entanglement across the interface is described by the same $c_{\text{eff}}$ \cite{ Wen2017}. For $c_{\text{LR}}$, which characterizes the energy transmission, it was shown in  
\cite{2020_Meineri} that $c_{\text{LR}}$ is independent of the details of initial state.

Although in some integrable examples these two quantities are related to each other~\cite{Eisler2010, Eisler_2012, brehm_2015_EE_interface_ising, Calabrese_2012_EE_junction}, in general they are independent~\cite{Bachas2020, Karch2024}. 
Remarkably, the authors in the recent work \cite{Karch2024} proposed the following inequalities:
\be 
\label{Eq:bound}
0\le c_{\text{LR}} \le c_{\text{eff}}\le \min\{c_{\text{L}}, \, c_{\text{R}}\},
\ee
where $c_\text{L}$ and $c_{\text{R}}$ are the central charges of CFTs on the left and right sides of the interface, respectively.
When the two CFTs are the same, we have $c_\text{L}=c_\text{R}=c$ and therefore $\min\{c_{\text{L}}, \, c_{\text{R}}\}=c$.
\eqref{Eq:bound} has the important implication that the amount of energy transmission can never exceed the amount of information transmission. This was verified based on the AdS/CFT correspondence and also in certain examples beyond holography \cite{Karch2024}.
For our later purpose, it is noted that the last inequality in \eqref{Eq:bound} indicates that the presence of a conformal interface will in general \textit{suppress} both the information transmission and the energy transmission. When $c_\text{L}=c_\text{R}=c$, the suppressions are characterized by $c_{\text{eff}}/c\in [0,1]$ and 
$c_{\text{LR}}/c\in [0,1]$, respectively.

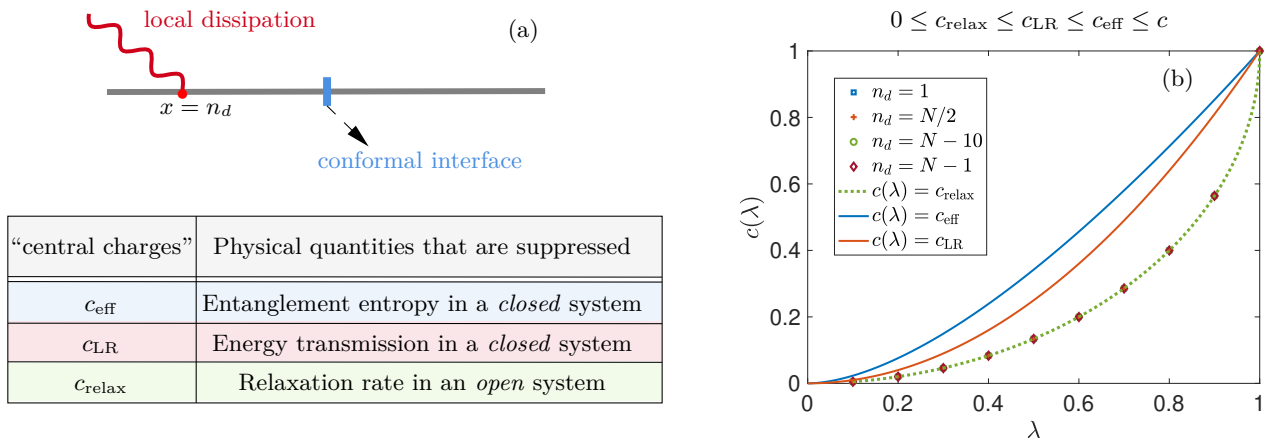
\begin{figure*}
\begin{tikzpicture}[x=0.75pt,y=0.75pt,yscale=-1,xscale=1]

\draw [color={rgb, 255:red, 128; green, 128; blue, 128 }  ,draw opacity=1 ][line width=2.25]    (100,101) -- (321,100.5) ;
\draw [color={rgb, 255:red, 74; green, 144; blue, 226 }  ,draw opacity=1 ][line width=3]    (211,93.5) -- (211,108) ;
\draw  [dash pattern={on 4.5pt off 4.5pt}]  (211,108) -- (230.57,127.1) ;
\draw [shift={(232,128.5)}, rotate = 224.31] [fill={rgb, 255:red, 0; green, 0; blue, 0 }  ][line width=0.08]  [draw opacity=0] (12,-3) -- (0,0) -- (12,3) -- cycle    ;

\draw  [color={rgb, 255:red, 208; green, 2; blue, 27 }  ,draw opacity=1 ][line width=1.5]  (90.89,61.53) .. controls (90.7,64.06) and (90.52,66.46) .. (91.94,67.57) .. controls (93.37,68.69) and (96.01,68.2) .. (98.78,67.68) .. controls (101.55,67.17) and (104.19,66.68) .. (105.62,67.79) .. controls (107.04,68.9) and (106.86,71.31) .. (106.67,73.83) .. controls (106.47,76.35) and (106.29,78.76) .. (107.72,79.87) .. controls (109.14,80.98) and (111.78,80.5) .. (114.55,79.98) .. controls (117.32,79.46) and (119.96,78.98) .. (121.39,80.09) .. controls (122.82,81.2) and (122.63,83.61) .. (122.44,86.13) .. controls (122.24,88.65) and (122.06,91.06) .. (123.49,92.17) .. controls (124.92,93.28) and (127.56,92.79) .. (130.33,92.28) .. controls (133.09,91.76) and (135.73,91.27) .. (137.16,92.39) .. controls (138.59,93.5) and (138.41,95.9) .. (138.21,98.42) .. controls (138.16,99.07) and (138.11,99.72) .. (138.09,100.33) ;

\draw (162,65) node  {$\textcolor[rgb]{0.82,0.01,0.11}{\text{local dissipation}}$};
\draw (205,130) node [anchor=north west][inner sep=0.75pt]  [color={rgb, 255:red, 74; green, 144; blue, 226 }  ,opacity=1 ]  {$\textcolor[rgb]{0.29,0.56,0.89}{\text{conformal interface}}$};

\draw (310,70) node  {(a)};

\draw (138.1,102.5) node  {$\textcolor{red}{\bullet}$};

\draw (145,110) node  {$x=n_d$};

            \node[inner sep=0pt] (russell) at (420pt, 130pt)
    {\includegraphics[width=.42\textwidth]{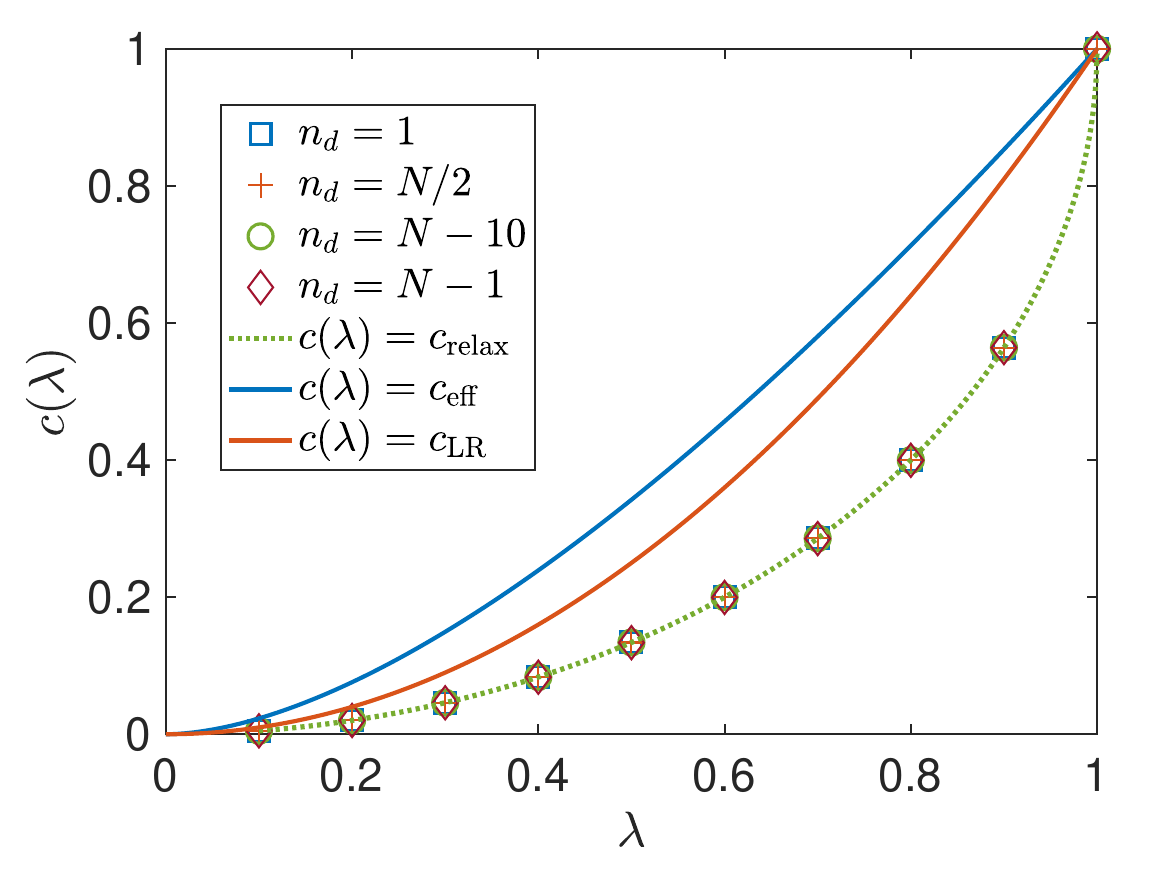}};
    
\draw (640,95) node  {(b)};

\draw (565,65) node {\textcolor{black}{$0\le c_{\text{relax}}\le c_{\text{LR}} \le c_{\text{eff}}\le c$}};

\begin{scope}[xshift=-45pt, yshift=95pt]
\draw  [draw opacity=0][fill={rgb, 255:red, 155; green, 155; blue, 155 }  ,fill opacity=0.1 ] (110,37) -- (440,37) -- (440,67) -- (110,67) -- cycle ;

\draw  [draw opacity=0][fill={rgb, 255:red, 74; green, 144; blue, 226 }  ,fill opacity=0.1 ] (110,70) -- (440,70) -- (440,90) -- (110,90) -- cycle ;

\draw  [draw opacity=0][fill={rgb, 255:red, 208; green, 2; blue, 27 }  ,fill opacity=0.1 ] (110,90) -- (440,90) -- (440,110) -- (110,110) -- cycle ;

\draw  [draw opacity=0][fill={rgb, 255:red, 126; green, 211; blue, 33 }  ,fill opacity=0.1 ] (110,111) -- (440,111) -- (440,130) -- (110,130) -- cycle ;

\draw   (110,37) -- (440,37) -- (440,131.25) -- (110,131.25) -- cycle ;
\draw    (110,67.5) -- (440,67.75) ;
\draw    (110,70) -- (440,70.25) ;
\draw    (110,91) -- (440,91.25) ;
\draw    (110,110.5) -- (440,110.75) ;
\draw    (205,37) -- (205,131.25) ;

\small
\draw (157.5,53) node  {``central charges''};

\draw (157,83) node  {$c_{\rm{eff}}$};
\draw (157,103) node  {$c_{\rm{LR}}$};
\draw (157,123) node  {$c_{\rm{relax}}$};

\draw (319,53) node  {Physical quantities that are suppressed};
\draw (319,82) node  {Entanglement entropy in a \textit{closed} system};
\draw (319,102) node  {Energy transmission in a \textit{closed} system};
\draw (319,122) node  {Relaxation rate in an \textit{open} system};
\end{scope}

\end{tikzpicture}
\caption{
Summary of the main results in this work.
(a) A sketch for introducing a local dissipation to a one-dimensional quantum critical system in the presence of a conformal interface. 
The total system is of length $L=2N$, the conformal interface is located at $x=N$, and the dissipation is introduced at $x=n_d$. 
(b) The relaxation coefficient $c_{\text{relax}}$ and its comparison with $c_{\text{LR}}$ and $c_{\text{eff}}$,
plotted as a function of the interface parameter $\lambda$ [see \eqref{eq:H2}] in a free-fermion lattice model at the critical point. The markers represent the exact numerical results of $c_{\text{relax}}$ for various choices of locations $n_d$ where the dissipation is introduced (Note: markers appear on the top of each other in the figure). 
For example, $n_d=1$ corresponds to the left boundary, and $n_d=N-1$ corresponds to the site next to the conformal interface.
The green dashed line is the analytical result of $c_{\text{relax}}$ in Eq.~\eqref{eq:crelax_analytic}, obtained from perturbative calculations. The blue and red lines are analytic results of $c_{\text{LR}}$ and $c_{\text{eff}}$ in Eqs.~\eqref{eq:cLR_energy} and \eqref{eq:ceff_entangle} respectively, with the corresponding lattice model calculations given in the appendix. Here the dissipation strength is $\gamma = 0.05$, and the central charge of the bulk critical theory is $c = 1$.
Bottom left: A summary of defining properties for 
$c_{\text{eff}}$, $c_{\text{LR}}$, and $c_{\text{relax}}$.
}
\label{fig:MainResult}
\end{figure*}

\subsection{Our motivation}

In this work, instead of considering a closed quantum critical system, we are interested in open systems by coupling the system to environment. It is known that even for a local dissipation, the system could finally reach a steady state \cite{Prosen_2008, znidaric2011_transpot_xxx, Kehrein2014_cubic_diss, znidaric2015_dissipation_relax, Prosen2015_BoundaryDrive, Shibata2020, Yamanaka2021, Tarantelli2021, chenshu2023}.

Now let us consider a one dimensional system with a conformal interface in the middle (see Fig.\ref{fig:MainResult}), and a local dissipation is introduced on the left side of the interface. If the conformal interface is totally transparent, i.e., there is essentially no defect, then the system will reach a steady state as usual. However, if the conformal interface is totally reflective, i.e., the system is cut into two halves that are decoupled from each other, then the right half will stay in the initial state forever and never reach the steady state. That is, the relaxation time of the total system will go to infinity. Equivalently, the relaxation rate, which is the inverse of relaxation time, is suppressed to be zero.
Now, as we tune the interface continuously from totally reflective to totally transmissive, it is expected that the 
interface will suppress the relaxation rates in a continuous way.

With the above intuitive picture in mind, we ask the following questions:

$\bullet$ For a quantum critical system with local dissipation,
how does the conformal interface suppress the relaxation rate? 

$\bullet$ Is this suppression described by one of the two quantities $c_{\text{eff}}$ and $c_{\text{LR}}$? If not, is it
described by any universal quantity?

For the second question, since $c_{\text{eff}}$ and $c_{\text{LR}}$ characterize the suppression of entanglement and energy transmission by the conformal interface in a \textit{closed} system, it is really an open question to us whether they are related to the suppression of relaxation rates in an \textit{open} system, where the quantum dynamics becomes non-untiary.
These basic questions motivate us to study a dissipative 
quantum critical system in the presence of a conformal interface.

\subsection{Main results}

To answer the above two questions, in this work we consider a Dirac free-fermion lattice at the critical point. The reason we consider this model is twofold: First, the conformal interface in this lattice model has been well studied when the system is closed \cite{Eisler2010, Eisler_2012, Wen2017, Eisler2022}, and one can extract both $c_{\text{eff}}$ and $c_{\text{LR}}$ from lattice calculations, which agree with the field theory predictions perfectly, as shown in our appendices.
Second, the Lindblad equation of an open quadratic fermionic or bosonic system can be explicitly solved by following the scheme of third quantization \cite{Prosen_2008, Prosen_2010_1, Prosen_2010_2}.\footnote{There are other free lattice systems that have these two nice properties, such as the  critical Ising model\cite{oshikawa1997_Ising_defect} and harmonic chain \cite{2012_Peschel_Eisler}, which deserve future studies.}

In the absence of the conformal interface, it is known that boundary dissipation in one dimensional integrable models generally leads to a universal cubic scaling of the Liouvillian gap (the relaxation rate of the dissipative system)~\cite{Baumgartner2008_semigroup_lindblad} with the system size $L$ as~\cite{Prosen_2008, znidaric2011_transpot_xxx, Kehrein2014_cubic_diss, znidaric2015_dissipation_relax, Prosen2015_BoundaryDrive, Shibata2020, Yamanaka2021, Tarantelli2021, chenshu2023} 
\be 
\label{eq:intro_gap_cubic} 
g \propto \frac{1}{L^{3}}.  
\ee 
The inverse $g^{-1}$ determines the relaxation time of the open system. Now, we introduce a conformal interface in the middle of the system (see Fig.\ref{fig:MainResult}),
parametrized by $\lambda\in [0,1]$, with $\lambda=0$ corresponding to a totally reflective interface, and $\lambda=1$ corresponding to a totally transmissive interface. For $\lambda\in(0,1)$, the conformal interface is partially transmissive and partially reflective. Interestingly, we find that the Liouvillian gap $g(\lambda)$ has the same scaling behavior as \eqref{eq:intro_gap_cubic}, i.e.,
\be 
\label{eq:intro_gap_cubic_interface} 
g(\lambda) \propto \frac{1}{L^{3}}, \quad \lambda\in [0,1].  
\ee 
The difference between \eqref{eq:intro_gap_cubic_interface} and \eqref{eq:intro_gap_cubic} are the prefactors in front of $1/L^3$, which characterize how the relaxation rate is suppressed by the conformal interface.
One key result we obtain in this work is that the suppression of relaxation rates is characterized by the following universal quantity:
\be
\label{eq:cdecay}
\dfrac{c_{\text{relax}}}{c} := \dfrac{g(\lambda)}{g(\lambda=1)} \approx 1- \sqrt{1-\lambda^{2}}\in [0,1].
\ee 
Here we define $c_{\text{relax}}$ as the relaxation coefficient, and $c=1$ is the central charge of the defect-free Dirac fermion CFT at $\lambda=1$. As expected, for $\lambda=0$ the relaxation rate is suppressed to be zero, and for $\lambda=1$ there is no suppression at all since the defect is totally transparent.

The result in \eqref{eq:cdecay} is obtained by two different approaches: One is an analytical perturbation calculation in the weak dissipation regime, i.e., when the dissipation strength is much smaller than the hopping strength (or equivalently the bandwidth of energy), which is the only energy scale in our lattice fermion model. The other is an exact calculation with numerics by diagonalizing the Liouvillian superoperator. See Fig.\ref{fig:MainResult} for the comparison of results from these two methods.

\medskip
We find that $c_{\text{relax}}$ is universal in the sense that:
\begin{enumerate}
\item  $c_{\text{relax}}$ is independent of the strength of dissipation in the weak dissipation regime.

\item $c_{\text{relax}}$ is independent of the location where we introduce the dissipation.

\end{enumerate}
In addition, if the local dissipation is introduced at the boundary, we find that $c_{\text{relax}}$ in \eqref{eq:cdecay} holds even when the dissipation strength is out of the weak 
dissipation regime (see section \ref{sec:finite}). 

By comparing with $c_{\text{eff}}$ and $c_{\text{LR}}$ in the same lattice model, we find that 
\be
\label{eq:MainResult}
0\le c_{\text{relax}}\le c_{\text{LR}} \le c_{\text{eff}}\le c,
\ee
where $c=1$ in our case, and the equalities hold only when the interface is totally reflective ($\lambda=0$) or totally transparent ($\lambda=1$), as shown in Fig.\ref{fig:MainResult}.
The distinct feature of $c_{\text{relax}}$ from 
$c_{\text{eff}}$ and $c_{\text{LR}}$ indicates that it is a new quantity to characterize the conformal interface in an open quantum critical system.

\medskip
The structure of the rest of this work is organized as follows. 
In Section~\ref{sec:model}, we introduce the setup of adding local dissipation to a critical system with a conformal interface, along with a review of known results on energy and entanglement transmission in a closed quantum critical system. 
In Section~\ref{sec:method}, we analyze the spectral structure of the Liouvillian using the third quantization formalism, by translating the problem to the study of an effective non-Hermitian Hamiltonian. Based on this, in Section~\ref{sec:gap}, we perform a perturbative analysis of the Liouvillian gap, which is the main focus of this work. In Sec.~\ref{sec:finite}, we analytically compute the Liouvillian gap for finite dissipation strength using a plane-wave ansatz, assuming that local dissipation is introduced at the boundary. By combining analytical results with numerical calculations, we demonstrate a universal suppression of relaxation rate, with results presented in Section~\ref{sec:suppress}. This allows us to define the relaxation coefficient $c_{\text{relax}}$ to characterize the conformal interface in a dissipative system. In Section~\ref{sec:discuss}, we conclude with a discussion and mention several future directions.
There are also several appendices: 
We show how to extract $c_{\text{LR}}$ and $c_{\text{eff}}$ from the free fermion lattice models in Appendix~\ref{Appendix:c_LR} and Appendix~\ref{Appendix:c_eff}, respectively.
Then in Appendix~\ref{app:gap_general_loc}, we give details on the perturbative calculation of the Liouvillian gap for arbitrary choices of the local dissipation location, as a supplement to the cases discussed in the main text. Finally, in Appendix~\ref{app:dynamics}, we study the out-of-equilibrium quantum dynamics of the particle density, showing how the dissipative system approaches its steady state.

\section{Setup and model} 
\label{sec:model}

\begin{figure}[htp]
    \centering
\begin{tikzpicture}
\node[inner sep=0pt] (russell) at (0pt,0pt)
    {\includegraphics[width=1.02\linewidth]{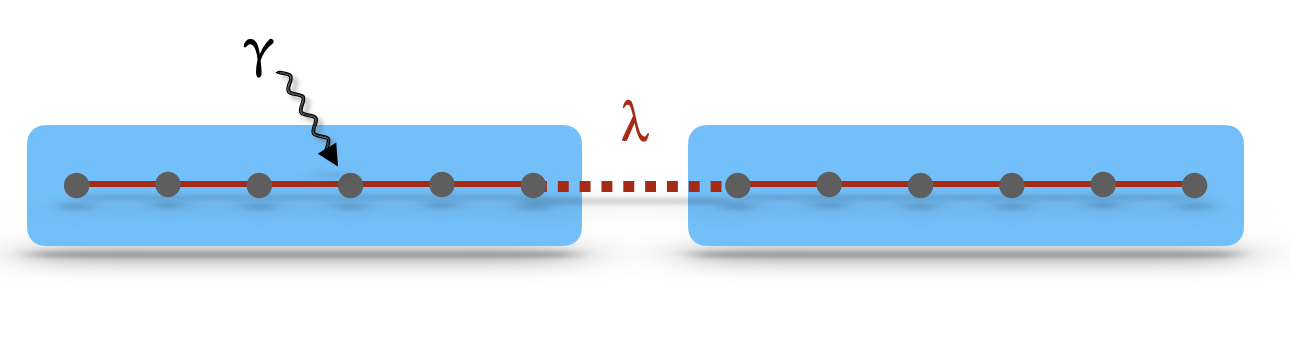}};

    \node at (-115pt, -22pt){$x=1$};
        \node at (-85pt, -22pt){$\cdots$};
    \node at (-55pt, -23pt){$n_d$};
        \node at (-35pt, -22pt){$\cdots$};
    \node at (-22pt, -22pt){$N$};
    \node at (16pt, -22pt){$N+1$};
            \node at (60pt, -22pt){$\cdots$};
            \node at (75pt, -22pt){$\cdots$};
        \node at (108pt, -22pt){$2N$};
\end{tikzpicture}
    \caption{Two free-fermion chains of size $N$ are connected via a conformal interface characterized by a parameter $\lambda$. A local dissipation of strength $\gamma$ is introduced at site $n_d \in [1, N]$ on the left chain. The total system size is $L=2N$.}
    \label{fig:lattice_model}
\end{figure}
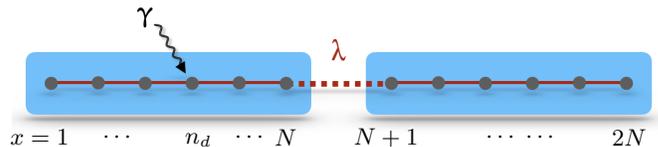

In this section, we introduce our setup by considering a dissipative quantum critical free fermionic chain with a conformal interface, where the dissipation is introduced locally at a single site. As show in Fig.~\ref{fig:lattice_model}, we begin with two identical free fermionic chains (each of length $N$) at criticality, which are coupled through a conformal interface~\cite{Eisler2010, Eisler_2012, Wen2017, Eisler2022}. The system is described by the following Hamiltonian
\be
\label{eq:H1}
H = \frac{J}{2}\sum\limits^{L=2N}_{m,n=1}H_{m,n}c^{\dagger}_{m}c_{n},
\ee
where $c_m^\dag$ and $c_n$ are fermionic operators that satisfy $\{c_m^\dag,c_n\}=\delta_{mn}$ and $\{c_m,c_n\}=\{c_m^\dag,c_n^\dag\}=0$.
$J$ is used to characterize the hopping amplitude in the system, and it is introduced for the purpose of comparing with the dissipation strength, as will be discussed later in Sec.\ref{sec:gap}. Throughout the whole work, we will take $J=1$.
The non-zero elements of the Hamiltonian matrix in \eqref{eq:H1} are 
\be
\begin{split}
\label{eq:H2}
H_{m,m+1} = H_{m+1,m} = \begin{cases}
    -1, \quad  m\neq N\\
    -\lambda, \quad m = N
\end{cases}\\
H_{N,N} = -H_{N+1,N+1} = \sqrt{1-\lambda^{2}} . 
\end{split}
\ee
Here the parameter $\lambda \in [0, 1]$ characterizes the conformal interface, with $\lambda = 0$ corresponding to a totally reflective interface, i.e., the two half chains are decoupled from each other, and $\lambda = 1$ corresponding to a
totally transmissive interface, i.e., the total system is a uniform chain of length $L=2N$. It is emphasized that we always have open boundary conditions at the two ends, as shown in 
Fig.~\ref{fig:lattice_model}.

The interface Hamiltonian in \eqref{eq:H1} is related to the defect-free case ($\lambda=1$) via a unitary transformation~\cite{Eisler2010, Eisler_2012, Eisler2022}, under which the energy spectrum is unchanged
\be 
\omega'_k = \omega_k = -\cos\left(\dfrac{k\pi}{2N+1}\right). 
\ee
Here we denote the energy spectrum for the interface Hamiltonian and the defect-free Hamiltonian as $\omega'_k$ and $\omega_k$, and the corresponding eigenfunctions will be denoted as $\phi'_k$ and $\phi_k$, respectively.
One can find that $\phi'_k$ and $\phi_k$ are related as
\begin{align}
\phi'_{k}(n) = 
    \begin{cases}
        \alpha_{k}\,\phi_{k}(n), \quad   n \in  [1, N],\\
        \beta_{k}\,\phi_{k}(n), \quad  n \in  [N+1, 2N],
    \end{cases} 
\label{eq:efunc}
\end{align}
where
\be 
\phi_{k}(n) = \sqrt{\frac{2}{2N+1}}\sin\left(\frac{k\pi n}{2N+1}\right), 
\ee
and the quasi-momentum $k = 1, 2, \dots, 2N$. 
The coefficients $\alpha_k$ and $\beta_k$ in \eqref{eq:efunc} take different values on the left and right side of the interface as
\be
\label{eq:alpha}
\alpha^{2}_{k} = 1 + (-1)^{k}\sqrt{1-\lambda^{2}}, \quad \beta^{2}_{k} = 1 - (-1)^{k}\sqrt{1-\lambda^{2}} ,
\ee
which reduce to unity in the defect-free limit $\lambda= 1$, as expected. 

Till now, we have not yet introduced the dissipation. The above setup is already enough to give the transmission coefficient~\cite{Eisler2010}
\be \label{eq:cLR_energy}
 c_{\text{LR}} =  \lambda^2,
\ee
as well as the effective central charge \cite{Eisler2010, Eisler_2012, Wen2017, Eisler2022}:
\be
\begin{aligned} \label{eq:ceff_entangle}
c_{\text{eff}} = -\frac{6}{\pi^2} \sum_{\kappa = \pm 1}
\Big\{ (1+\kappa \lambda) \log (1+ \kappa \lambda) \log \lambda \\ 
+ (1+\kappa \lambda) {\rm Li}_2(-\kappa \lambda)  \Big\} , 
\end{aligned}
\ee
where $\lambda\in[0,1]$.
See Fig.\ref{fig:MainResult} for a comparison of $c_{\text{LR}}$ and $c_{\text{eff}}$. It can be observed that $c_{\text{LR}}(\lambda)\le c_{\text{eff}}(\lambda)$, which is conjectured to be true for general CFTs \cite{Karch2024}.
See also Appendix~\ref{Appendix:c_LR} and Appendix~\ref{Appendix:c_eff} for how to extract $c_{\text{LR}}$ and $c_{\text{eff}}$ from the lattice models by studying directly the energy transmission and entanglement transmission.

\medskip

Now we introduce a local dissipation of strength $\gamma$, which is applied to a single site at $x=n_d$ in the left half chain, as shown in Fig.~\ref{fig:lattice_model}. The dissipation we consider is simply a particle loss/gain to remove/add fermions from the system, which makes the system out of equilibrium. 
The dynamics of the system can be described by Lindblad master equation~\cite{breuer2002theory, manzano2020_intro_lindblad} 
\be
\dfrac{d\rho}{dt} = \mathcal{\hat{L}}[\rho]
 = -i[H, \rho] + \sum_{\mu=\pm}\left(L_{\mu}\rho L^{\dagger}_{\mu} - \frac{1}{2}\{L^{\dagger}_{\mu}L_{\mu}, \rho\}\right)
 \label{eq:lindblad} ,
\ee 
where $\rho(t)$ is the density matrix of the system at time $t$, 
and $\mathcal{\hat{L}}$ is called
the \textit{Liouvillian} (also called \textit{Lindbladian}) superoperator that generates the dissipative dynamics. 
Here $H$ is the Hamiltonian in Eq.(\ref{eq:H1}) and(\ref{eq:H2}) that drives the unitary dynamics.
For the jump operators $L_\mu$, we consider the following choice : 
\be 
\label{eq:jump_op}
L_+ = \sqrt{\gamma_{+}}c_{n_d}^\dagger, \quad L_- = \sqrt{\gamma_{-}}c_{n_d},
\ee
which describe the local particle gain and loss at site ${n_d} \in [1, N]$, and the dissipation strengths $\gamma_+$ and $\gamma_-$ are non-negative, i.e., $\gamma_{\pm} \ge 0$, such that the positivity of $\rho(t)$ is preserved. 
For later purposes, we will define the total dissipation strength $\gamma$ as 
\be
\label{eq:Gamma}
\gamma=\gamma_+ +\gamma_-.
\ee
As we will see later, for the choice of jump operators in \eqref{eq:jump_op}, the relaxation time depends only on $\gamma$, and is independent of the concrete values of $\gamma_+$ and $\gamma_-$.

Since our jump operators in \eqref{eq:jump_op}
are linear in the fermionic operators, the resulting Liouvillian is quadratic~\cite{Prosen_2008, Prosen_2010_1,Prosen_2010_2, cirac2013_fermion_lindblad, yikang2021, Yamanaka2021, landi2023}. This allows us to solve the Liouvillian spectrum as well as the relaxation time numerically even for a large system size \cite{Prosen_2008, Prosen_2010_1, Prosen_2010_2}.

\section{Spectral structure of the Liouvillian: Third quantization formalism} \label{sec:method}

To analyze the spectral properties of the quadratic Liouvillian operator in \eqref{eq:lindblad}, we employ the third quantization formalism. 
We will firstly give a brief review of the general method as developed in \cite{Prosen_2008, Prosen_2010_1, Prosen_2010_2}, and then apply it to our setup.

As a first step in applying this method, we perform a basis transformation from complex Dirac fermion operators to Majorana fermion operators, under which the Hamiltonian $H$ in \eqref{eq:lindblad} is of the form
\be
\label{eq:maj}
H_{\text{majorana}} = \sum\limits_{j,k = 1}^{4N}w_{j}\text{H}'_{jk}w_{k} = \underline{w}\cdot\textbf{H}'\underline{w},
\ee
and the jump operators are 
\be
L_{\mu} =  \sum\limits_{j = 1}^{4N} l_{\mu , j}w_{j} =  \underline{l}_{\mu}\cdot \underline{w},  
\ee
where $w_{2j-1} = c_{j} + c^{\dagger}_{j}$ and $w_{2j} = i(c_{j} - c^{\dagger}_{j})$ are Majorana fermionic operators that satisfy the anti-commutation relations $\{w_j, w_k\} = 2\delta_{j,k}$. This leads to the definition of a Hermitian matrix $\textbf{M}$ that describes the dissipation effect, whose elements are given by
\be
\text{M}_{ij} = \sum_{\mu}l_{\mu,i}l^{\ast}_{\mu,j} .
\ee
It is noted that the information of dissipation strength $\gamma_\mu$ is encoded in the definition of $l_{\mu, j}$, where $\mu$ labels the channel. In our case $\mu =\pm$, as seen from \eqref{eq:jump_op}.

In the third quantization formalism, the Liouvillian $\mathcal{\hat{L}}$ is treated as a linear map over a space of operators of dimension $4^{2N}$, where each vector corresponds to an operator of $4N$ Majorana fermions. The inner product of this operator space is defined as $\langle A | B\rangle = 2^{-2N} \Tr (A^{\dagger}B)$, where $2^{-2N}$ is the normalization factor.  
In this operator space, the Liouvillian can be re-expressed in terms of $8N$ Hermitian Majorana fermionic maps $\hat{a}_{i}$ as~\footnote{It is reminded that here we have $8N$ maps instead of $4N$, because the original complex-fermion chain is of length $L=2N$ instead of $N$.} 
\be
\label{eq:structure_matrix_A}
    \mathcal{\hat{L}} = \hat{\underline{a}}\cdot\textbf{A}\hat{\underline{a}} - A_{0}\hat{\mathbb{I}}, 
\ee
where $\textbf{A}$ is called the structure matrix
\be
\label{eq:matrix_A}
\mathbf{A} = \begin{bmatrix}
    -2i\mathbf{H}' + i \Im[\mathbf{M}]  &&i\mathbf{M} \\
    -i\mathbf{M}^{T} && -2i\mathbf{H}' -i \Im[\mathbf{M}] 
\end{bmatrix} ,
\ee
and $A_0 = \Tr [ \textbf{M} ]$.
The structure matrix $\textbf{A}$ is complex anti-symmetric, and its eigenvalues occur in pairs of the form $\{\beta_j, - \beta_j\}$, where $\beta_j$ denotes an eigenvalue with a non-negative real part, known as the \textit{rapidities}. 
They determine the full spectrum of the Liouvillian as~\cite{Prosen_2010_1}
\be \label{eq:liouvillian_rapidity}
\Lambda_{i} = - 2\sum\limits^{4N}_{j=1}\beta_{j}v_{ji},
\ee 
where $j = 1, 2, \dots, 4N$ and $v_{ji}$ is an element of vector $v_{i}= (v_{1i}, v_{2i}, \dots, v_{4Ni})$, $v_{ji} \in \{0,1\}$ that forms a set of orthogonal basis of the operator space. 
For a finite system, the Liouvillian always has at least one zero eigenvalue (those eigenvalues with a non-vanishing real part should have a negative real component, ensuring that the corresponding modes decay with a finite lifetime), corresponding to the steady state~\cite{Baumgartner2008_semigroup_lindblad}. From Eq.~\eqref{eq:liouvillian_rapidity}, if the rapidity spectrum does not include eigenvalues with a vanishing real part, then the Liouvillian spectrum will contain only a single eigenvalue with a vanishing real part ($\Lambda = 0$), corresponding to the case where $v_{ji}$ is zero for all $i$, i.e., $v_i = (0, 0, \dots, 0)$. Otherwise, multiple eigenvalues with vanishing real parts will appear in the Liouvillian spectrum.

The structure matrix \textbf{A} can be unitarily transformed to a block-triangular matrix~\cite{Prosen_2010_1} 
\be \label{eq:structure_A_triangular}
\tilde{\textbf{A}} =\textbf{U}\textbf{A}\textbf{U}^{\dagger} = \begin{bmatrix} 
-\textbf{X}^{T} & 2i \Im[\textbf{M}]
\\
0 & \textbf{X}
\end{bmatrix}
\ee
where $\textbf{X} = -2i\textbf{H}' + \Re[\textbf{M}]$ and $\textbf{U}$ is a permutation transformation. Furthermore, the matrix $\textbf{X}$ can be transformed into the following one by a unitary transformation~\cite{Prosen_2010_1,Yamanaka2021} 
\be \label{eq:structure_A_similar_X}
\tilde{\textbf{X}} =\textbf{S}\textbf{X}\textbf{S}^{\dagger} = \begin{bmatrix} 
-i\textbf{Z}^{\dagger} & 0
\\
0 & i\textbf{Z}
\end{bmatrix},
\ee
where \textbf{S} is a unitary matrix and the matrix $\mathbf{Z}$ has the following expression
\be\label{eq:effective_Ham}
\textbf{Z} = \frac{1}{2}\left( \textbf{H} - i\frac{\mathbf{\Gamma}}{2}\right) ,
\ee
where $\textbf{H}$ is the Hamiltonian matrix in the original complex Dirac fermion representation, and $\mathbf{\Gamma}$ is a (diagonal) matrix that counts the local dissipation applied at each site of the system. 
For our case where the local dissipation is applied at site $n_d$ with the jump operators chosen in \eqref{eq:jump_op}, its elements are
\be \label{eq:dissipation_Gamma_matrix}
\mathbf{\Gamma}_{n,n'} = \gamma \, \delta_{n,n_d}\delta_{n',n_d} ,
\ee
which take a non-zero value only at site $n_d$ where the local dissipation is applied, with $\gamma$ defined in \eqref{eq:Gamma}.
Then, the eigenvalues of \textbf{A} 
in \eqref{eq:structure_matrix_A} and \eqref{eq:matrix_A} 
are reduced to that of a non-Hermitian matrix $\mathbf{Z}$, which has a simple relation to the Hamiltonian and the local dissipation.

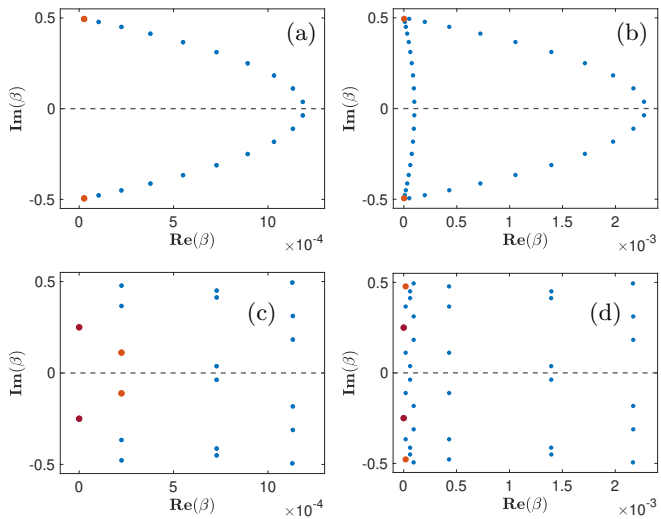
\begin{figure}
    \centering
    \begin{tikzpicture}
    \node[inner sep=0pt] (russell) at (-200pt,-85pt)
    {\includegraphics[width=.26\textwidth]{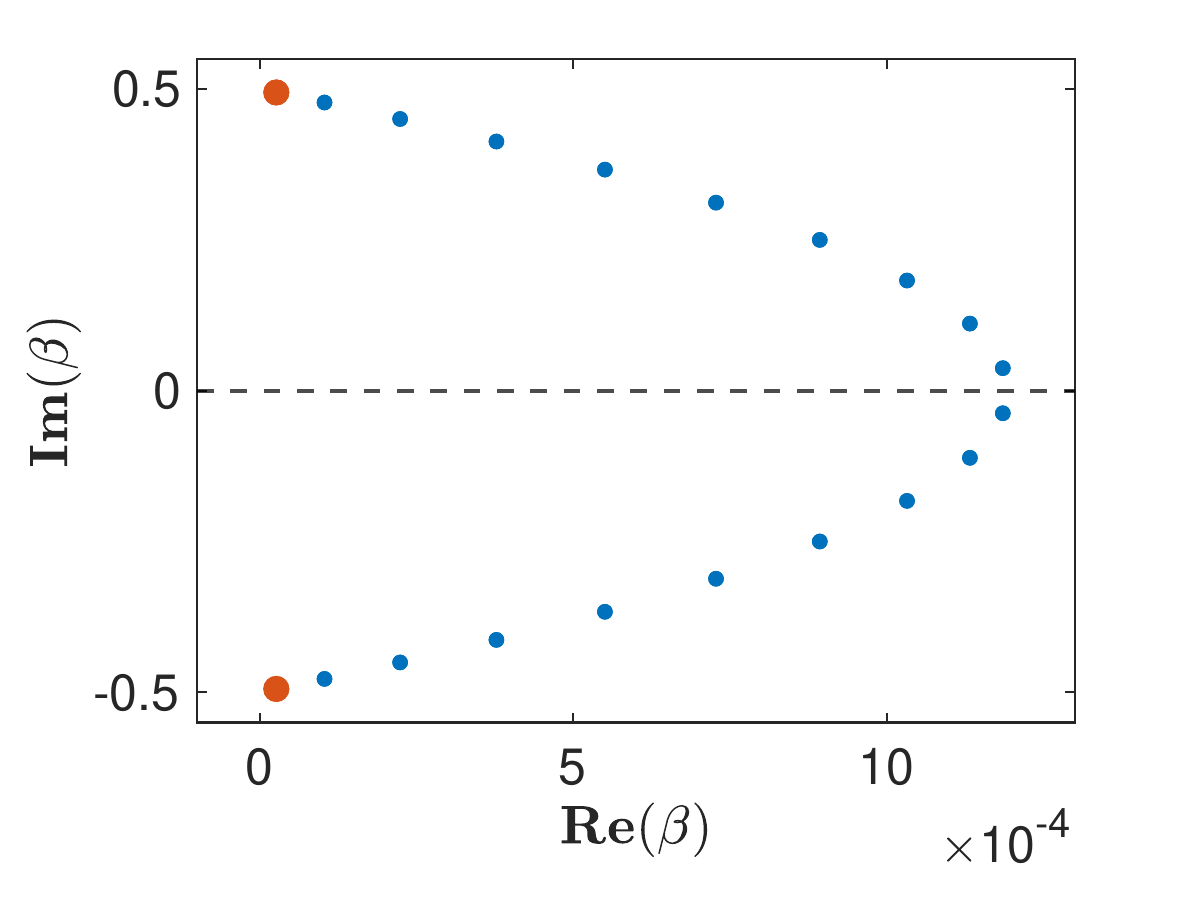}};
    \node[inner sep=0pt] (russell) at (-74pt,-85pt)
    {\includegraphics[width=.26\textwidth]{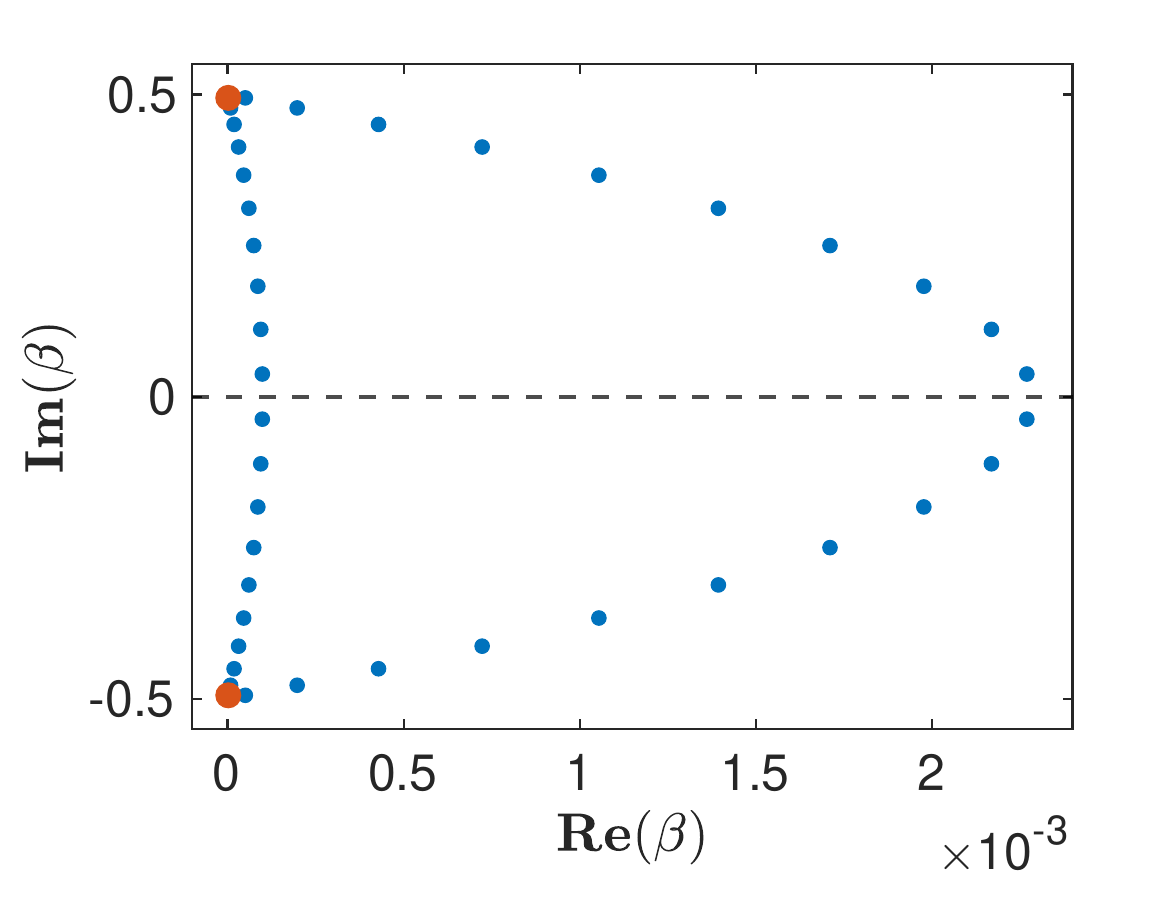}}; 
    \node[inner sep=0pt] (russell) at (-200pt,-185pt)
    {\includegraphics[width=.26\textwidth]{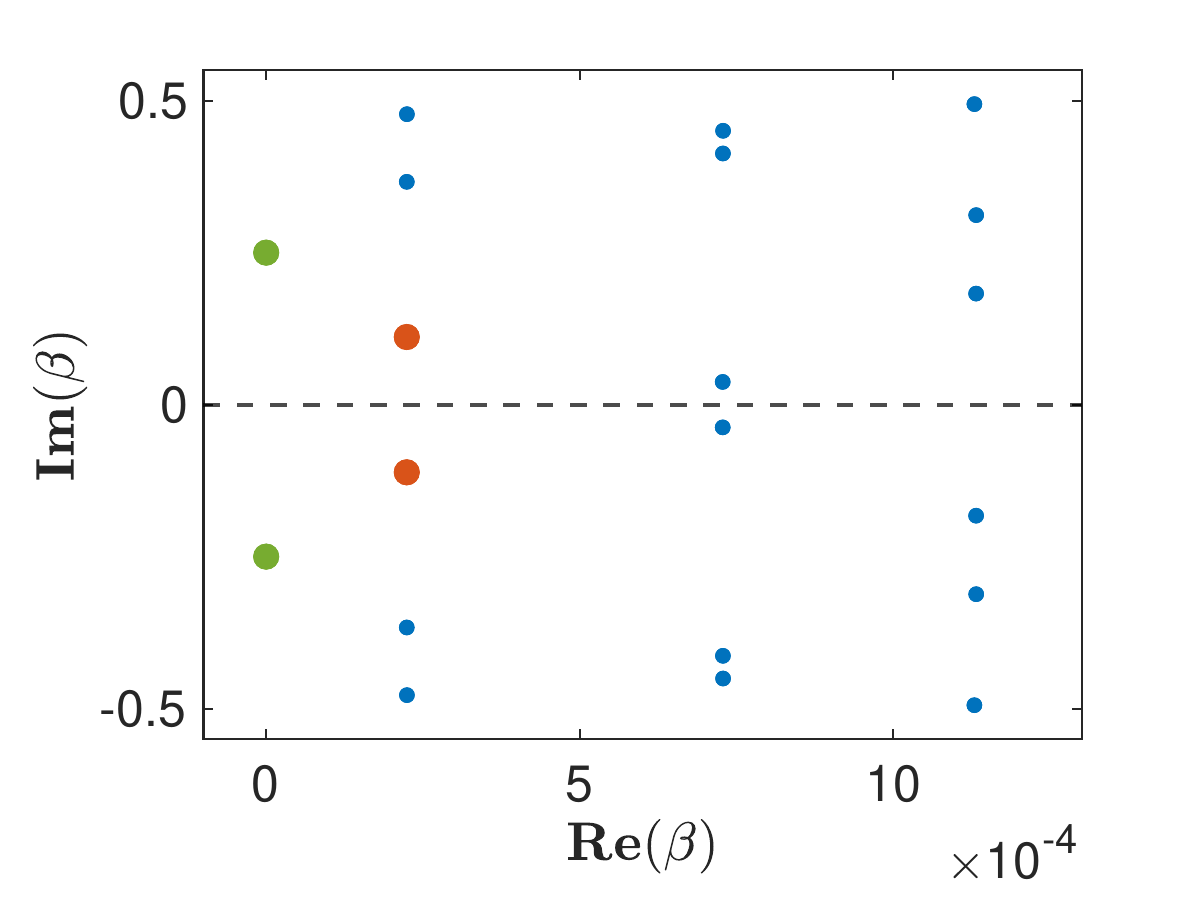}};
    \node[inner sep=0pt] (russell) at (-74pt,-185pt)
    {\includegraphics[width=.26\textwidth]{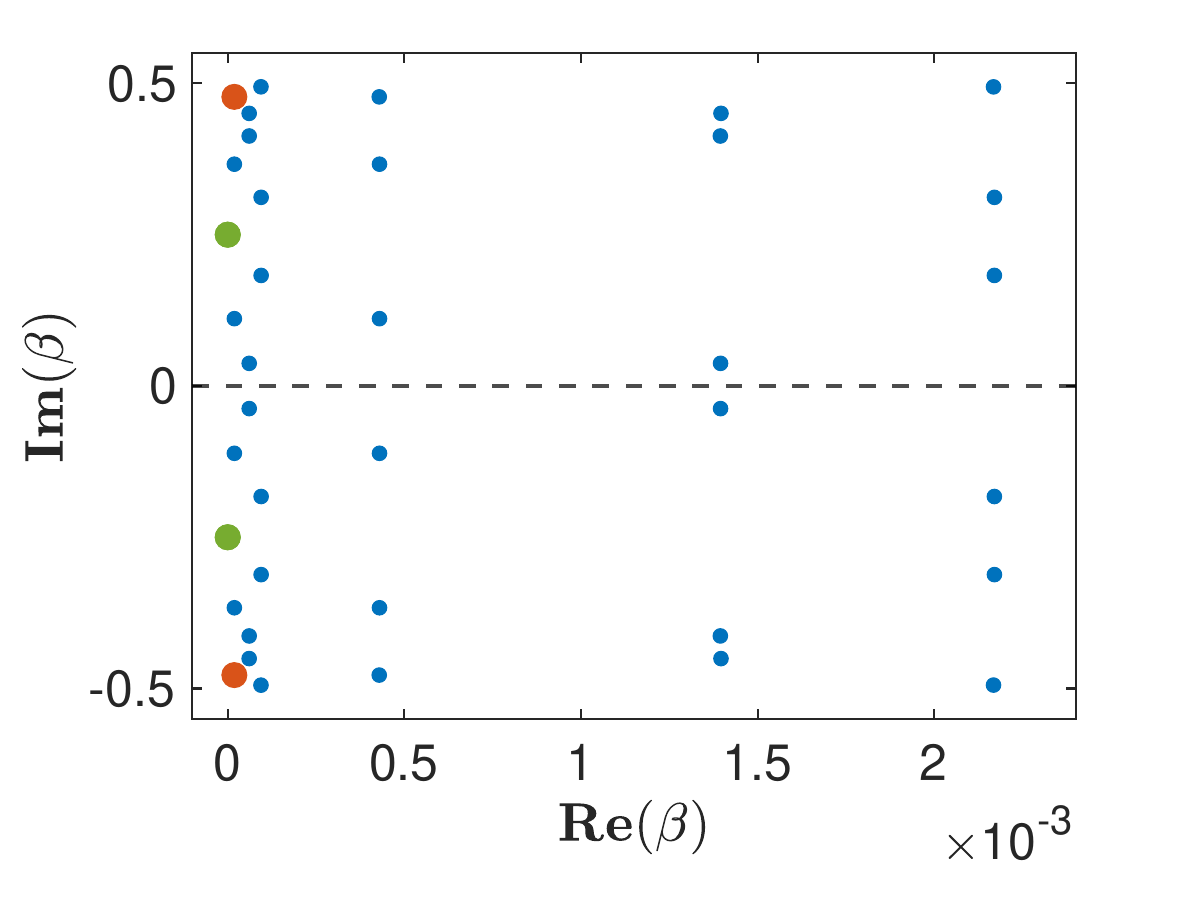}};
    \node at (-155pt, -52pt){(a)};
    \node at (-30pt, -52pt){(b)};
    \node at (-170pt, -158pt){(c)};
    \node at (-40pt, -158pt){(d)};
\end{tikzpicture}

\caption{Rapidity spectrum $\{\beta_i\}$ for the free fermionic chain of length $L=2N = 20$, with dissipation strength $\gamma = 0.05$. (a) and (b): The local dissipation is added on the left boundary, i.e. $n_d = 1$, and the interface parameter is set as (a) $\lambda = 1$ (totally transmissive) and (b) $\lambda = 0.4$ (partially transmissive). (c) and (d): The local dissipation is added right next to the interface, i.e. at $n_d = N-1$, and the interface parameter is set as (c) $\lambda = 1$ (totally transmissive), and (d) $\lambda = 0.4$ (partially transmissive). 
Orange dots represent rapidities with smallest real part other than zero, which determines the Liouvillian gap. 
Green dots represent rapidities with a vanishing real part and correspond to oscillating modes. Their appearance depends on the location of dissipation $n_d$ and the system size $2N$. All higher modes (decay modes) are denoted by blue dots.}
\label{fig:spec_at_edge}
\end{figure} 

Before applying the above procedure to our setup, let us briefly introduce several key properties and concepts about the Liouvillian spectrum $\{\Lambda_i\}$. First, all $\Lambda_i$ satisfy $\Re(\Lambda_i)\le 0$. Second, a non-equilibrium steady state is an eigenstate of the Liouvillian $\mathcal{\hat{L}}$  with eigenvalue $0$. This steady state is time-independent and free from decoherence.
Third, in certian cases, there are modes with nonzero purely imaginary eigenvalues, which is a phenomenon known as \textit{oscillatory coherence}~\cite{Baumgartner2008_semigroup_lindblad, Baumgartner2012_semigroup, liang2013_symm_lindblad}.
These modes are time-dependent but free from decay.
In this case, depending on the initial state, the system does not necessarily relax to a steady state. This is totally fine, since in this work we are mainly interested in the relaxation time, i.e, the longest lifetime of the decay modes (all the modes other than the steady state and the oscillating modes).

Next, by using the third quantization formalism introduced above, we study the spectral structure of the Liouvillian in our setup. 
In particular, we investigate the rapidity spectrum $\{ \beta_i \}$ that were introduced as the eigenvalues of the structure matrix $\textbf{A}$ in Eq.~\eqref{eq:structure_matrix_A}. 
From Eqs.~\eqref{eq:structure_A_triangular} and \eqref{eq:structure_A_similar_X}, the structure matrix $\textbf{A}$ and the matrix $\widetilde{\textbf{X}}$ share the same eigenvalues. Based on this, the rapidity $\{ \beta_i \}$ can be written in terms of the eigenvalues $\{E_i\}$ of the matrix $\textbf{Z}$ as 
\be \label{eq:relation_rapidity_Z}  
\beta_i = \{- i E_i^*, i E_i\}
\ee
where the conjugate pair comes from the diagonal structure shown in Eq.~\eqref{eq:structure_A_similar_X}, and we have chosen the ordering of $0 \le \Re[\beta_1] \le \Re[\beta_2] \le \cdots \le \Re[\beta_{4N}]$ and $0 \ge \Im[E_1] \ge \Im[E_2] \ge \cdots \ge \Im[E_{2N}]$.

Based on the above relation, we numerically calculate the rapidity spectrum $\{ \beta_i \}$ in our setup as introduced in Sec.\ref{sec:model}, with the sample plots shown in Fig.~\ref{fig:spec_at_edge}. We consider two different choices of $n_d$ where we introduce local dissipation: (i) the dissipation is introduced at the left boundary at $n_d=1$, and (ii) the dissipation is introduced at the site nearest to the conformal interface, at $n_d=N-1$. It is reminded that the conformal interface in the lattice model is located at sites $N$ and $N+1$, and the total length of the system is $L=2N$.

We study the above two cases with different values of interface parameters $\lambda \in [0, 1]$ which control the transmission property through the conformal interface. Several interesting features can be found in Fig.~\ref{fig:spec_at_edge}:

First, by fixing the dissipation location $n_d$, one can find that as we decrease the interface parameter $\lambda$, modes with smaller real part of eigenvalues will appear. Physically, this means that the relaxation rates of these modes are reduced, requiring more time for the system to reach a steady state. This is as expected, since a smaller $\lambda$ 
results in a smaller transmission coefficient, which makes it harder for the system to reach a steady state.

\begin{figure}[tp]
    \centering
\begin{tikzpicture}
\node[inner sep=0pt] (russell) at (-200pt,-85pt)
    {\includegraphics[width=.26\textwidth]{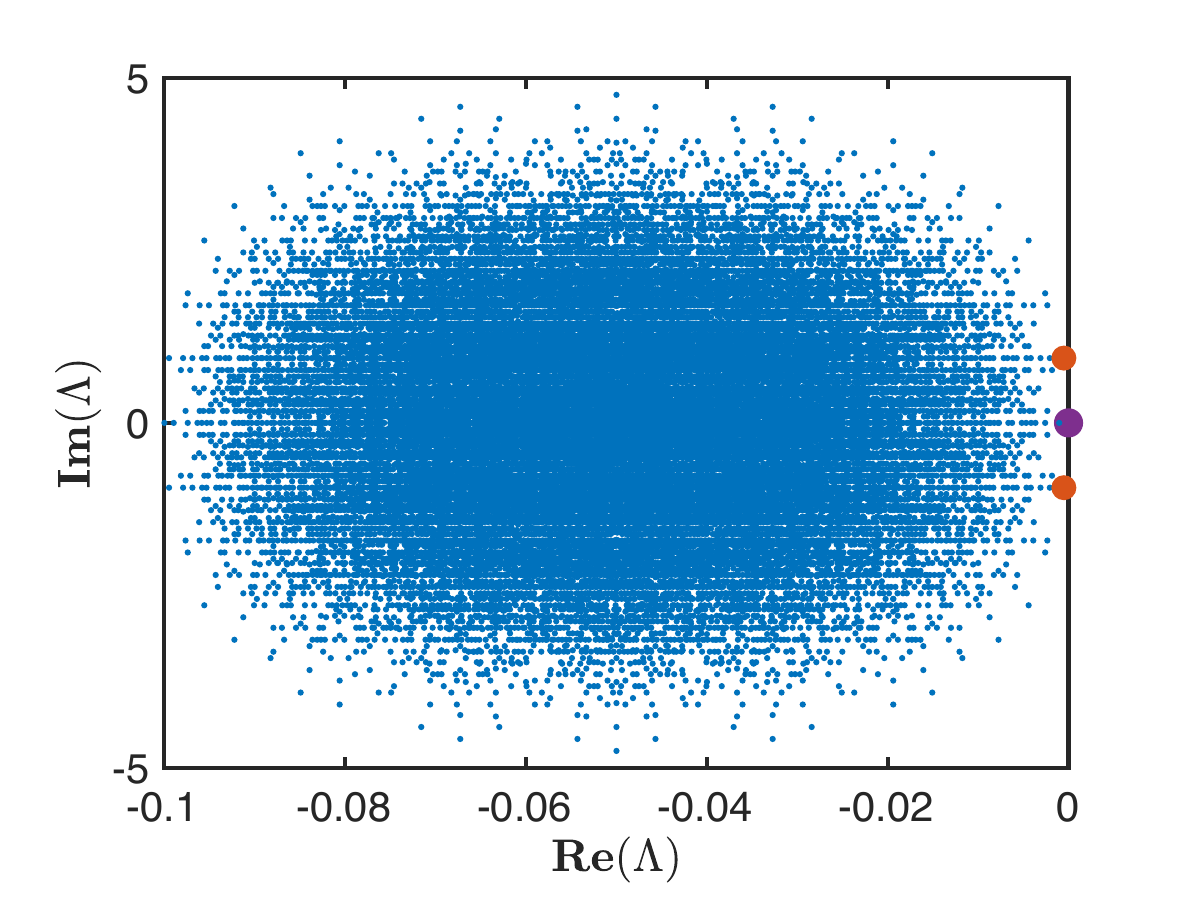}};
    \node[inner sep=0pt] (russell) at (-73pt,-85pt)
    {\includegraphics[width=.26\textwidth]{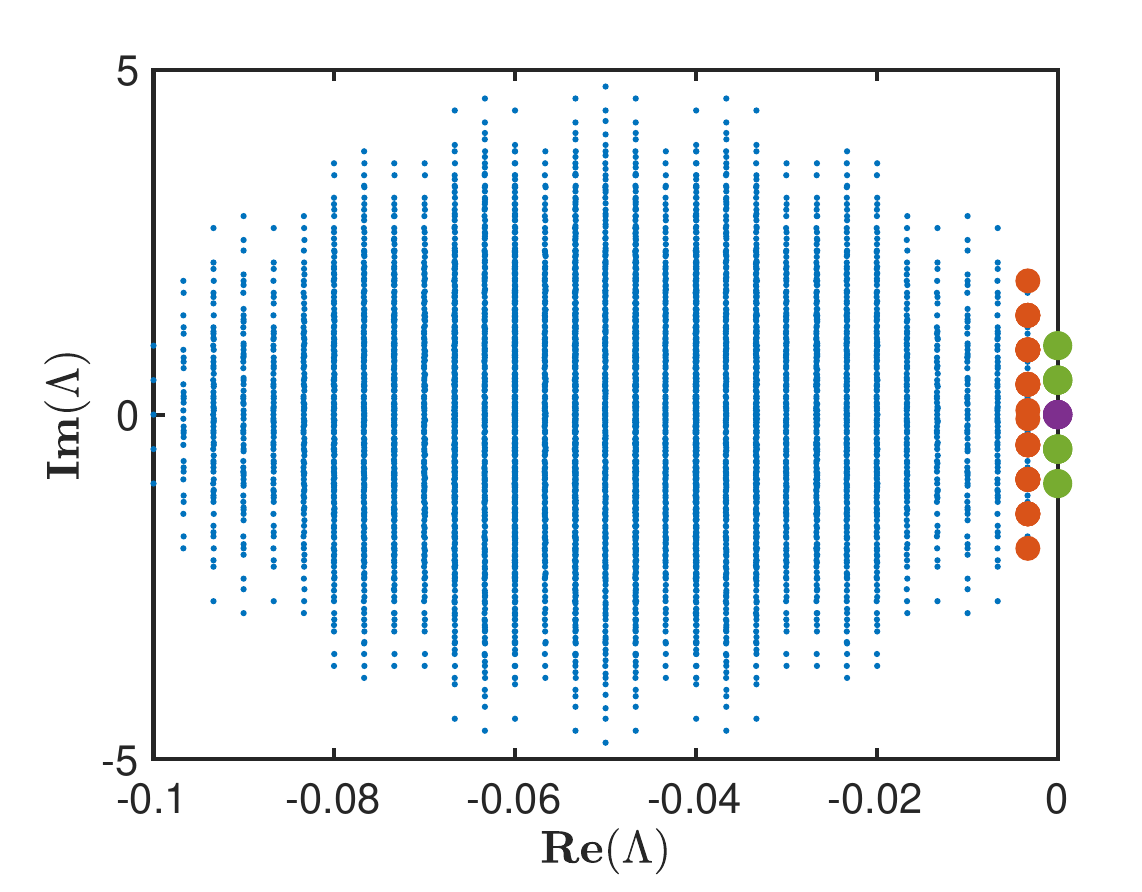}}; 
    \node at (-155pt, -46pt){(a)};
    \node at (-22pt, -46pt){(b)};
\end{tikzpicture}
    \caption{Full spectrum of the Liouvillian superoperator for a total size $L=2N=8$, with dissipation strength $\gamma = 0.05$, and $\lambda = 0.8$. The local dissipation is applied at (a) $n_d=1$ and (b) $n_d=N-1$.  All eigenvalues of the Liouvillian come in complex conjugate pairs with non-positive real parts. Purple dots show the steady-state solution. Green dots are the oscillatory modes with zero real part. Orange dots represent the slowest decaying modes, and blue dots represent all higher modes.}
    \label{fig:spectrum}
\end{figure}

Second, by fixing the interface parameter $\lambda$, 
as we change the location of dissipation $n_d$, the distribution of the rapidity spectrum $\{ \beta_i \}$ can change in a dramatic way. This indicates that the relaxation rate of the system strongly depends on the location of dissipation.  This strong dependence can be more clearly seen by looking at 
the full spectrum $\{\Lambda_i\}$ of the Liouvillian superoperator in Fig.\ref{fig:spectrum}, which 
are obtained based on the rapidity spectrum and Eq.~\eqref{eq:liouvillian_rapidity}.

The strong dependence of Liouvillian spectrum on the 
dissipation location $n_d$ gives rise to a 
natural question: Can we extract a universal quantity characterizing
the conformal interface, which is independent of the dissipation location
$n_d$? We will give a positive answer to this question in the next 
two sections, by studying the Liouvillian gap analytically.

\section{The Liouvillian gap: from perturbation theory to numerics}\label{sec:gap}

The main focus of this section is to investigate the influence of the conformal interface to the Liouvillian gap, i.e. the gap of the real part of the Liouvillian. 
Since the real part of Liouvillian spectrum is bounded by 0 from above, the Liouvillian gap is defined as
\be \label{eq:gap_Lambda}
g = - \max_{\Re[\Lambda_{i}] \neq 0} \left\{\Re[\Lambda_{i}] \right\}, 
\ee
the inverse of which corresponds to the relaxation time of the system~\footnote{We want to point out that there were recent examples where the Liouvillian gap does not determine the relaxation time, due to the diffusive transports in the dissipated many-body quantum system~\cite{mori2020_gap_relax}.
} .
From the relation in Eq.~\eqref{eq:liouvillian_rapidity}, the Liouvillian gap can be written as
\be
g = 2 \min_{\Re[\beta_{i}] \neq 0} \left\{ \Re[\beta_{i}]
\right\},
\ee
which corresponds to the case with a single non-zero value in the vector $v_i$ defined in Eq.~\eqref{eq:liouvillian_rapidity}. 
As we have already discussed in Sec.~\ref{sec:method}, the rapidity spectrum $\{ \beta_i \}$, which consists of the eigenvalues of the structure matrix $\textbf{A}$, can be derived from the non-Hermitian matrix $\textbf{Z}$ in \eqref{eq:effective_Ham}. Then, one can express the Liouvillian gap in the following form 
\be \label{eq:gap_def_Z}
g = - 2 \max_{\Im[E_{i}] \neq 0} \left\{\Im[E_{i}] \right\},
\ee 
where $E_i$ are the eigenvalues of $\textbf{Z}$.
That is, the problem is reduced to solving the eigenvalues ${ E_i }$ of the matrix $\textbf{Z}$ and identifying the largest real part of the eigenvalues other than zero. 
As shown in Eq.~\eqref{eq:effective_Ham}, this matrix $\textbf{Z}$ combines the Hamiltonian (in the complex Dirac fermion representation) and the dissipation term. In particular, we can consider it as an effective Hamiltonian of the dissipative dynamics, which has the following form
\be \label{eq:Heff_nonherm}
\begin{aligned}
& H_{\text{eff}} = \left( c_1^\dagger, \ \dots, \ c_{2N}^\dagger \right) \textbf{H}_{\text{eff}} \left( \begin{matrix}
c_1 \\ \vdots \\ c_{2N} 
\end{matrix} \right) 
\\ & = \left( c_1^\dagger, \ \dots, \ c_{2N}^\dagger \right) 2 \textbf{Z} \left( \begin{matrix}
c_1 \\ \vdots \\ c_{2N} 
\end{matrix} \right) 
= H - i \frac{\gamma}{2} c_{n_d}^\dagger c_{n_d} ,
\end{aligned}
\ee
where the imaginary onsite potential comes from the matrix $\mathbf{\Gamma}$ that counts the dissipation applied to each site, as discussed in Eq.~\eqref{eq:dissipation_Gamma_matrix}.
This effective Hamiltonian also governs the dynamics of the system in the single-particle sector~\cite{Prosen_2008, cirac2013_fermion_lindblad, Yamanaka2021, Alba2023} (see Appendix~\ref{app:dynamics} for more details).

Now we attempt to gain an analytical insight into the spectrum of the effective Hamiltonian in \eqref{eq:Heff_nonherm}. In the presence of local dissipation, there is no longer a well-defined (quasi)-momentum, making it difficult to obtain an exact solution. Nevertheless, based on the known results for the dissipation-free case, which has been discussed in Sec.~\ref{sec:model}, we are able to calculate the spectrum perturbatively when the dissipation
strength $\gamma$ is small comparing to the hopping strength $J$ in 
\eqref{eq:H1}, i.e., $\gamma \ll J$, where we take $J=1$ throughout the work. 
Similar to the treatment of boundary dissipation in Ref.~\cite{Shibata2020}, we consider the first-order perturbation. The change of energy spectrum of the effective Hamiltonian $H_{\text{eff}}$ can be obtained as:
\be
\begin{aligned}
\Delta \omega'_k & = \langle \phi'_k | - i \frac{\gamma}{2} c_{n_d}^\dagger c_{n_d} | \phi'_k \rangle + \mathcal{O}(\gamma^2) 
\\ & 
= - i \frac{\gamma}{2} \phi'^*_k(n_d) \phi'_k(n_d) + \mathcal{O}(\gamma^2), 
\end{aligned}
\ee
where $|\phi'_k\rangle$ is the eigenvector 
of the Hamiltonian in \eqref{eq:H1} and 
$\phi'_k(n)$ is given in \eqref{eq:efunc}.
Therefore, up to the first-order perturbation, we have the spectrum of the non-hermitian matrix $\textbf{Z} = \frac{1}{2} \textbf{H}_{\text{eff}}$ as
\be 
\begin{aligned}
E_k & = \frac{1}{2} \left[  \omega'_k + \Delta \omega'_k \right] 
\approx \frac{1}{2} \left[  \omega'_k - i \frac{\gamma}{2} \phi'^*_k(n_d) \phi'_k(n_d) \right] 
\\ & 
\approx - \frac{1}{2} \cos \left(\frac{k \pi}{2N+1}\right) - i \frac{\gamma}{2} \frac{\alpha_k^2}{2N+1} \sin^2 \left(\frac{n_d k \pi}{2N+1} \right),
\end{aligned}
\ee
where $n_d$ is location of the dissipation and the factor $\alpha_k$ is a function of both the interface parameter $\lambda$ and the (quasi-)momentum $k$ as shown in Eq.~\eqref{eq:alpha}.  
From Eq.~\eqref{eq:gap_def_Z}, we then have the following perturbative expression of the Liouvillian gap
\be\label{eq:Liouvillian_gap_perturbation}
g(\lambda) \approx \gamma \min \left\{ \frac{\alpha^{2}_{k}}{2N+1} \sin^2 \left(\frac{n_d k \pi}{2N+1}\right) \right\},
\ee
which is valid upto the first order of the dissipation strength $\gamma$.

\begin{figure}
    \centering
    \includegraphics[width=0.8\linewidth]{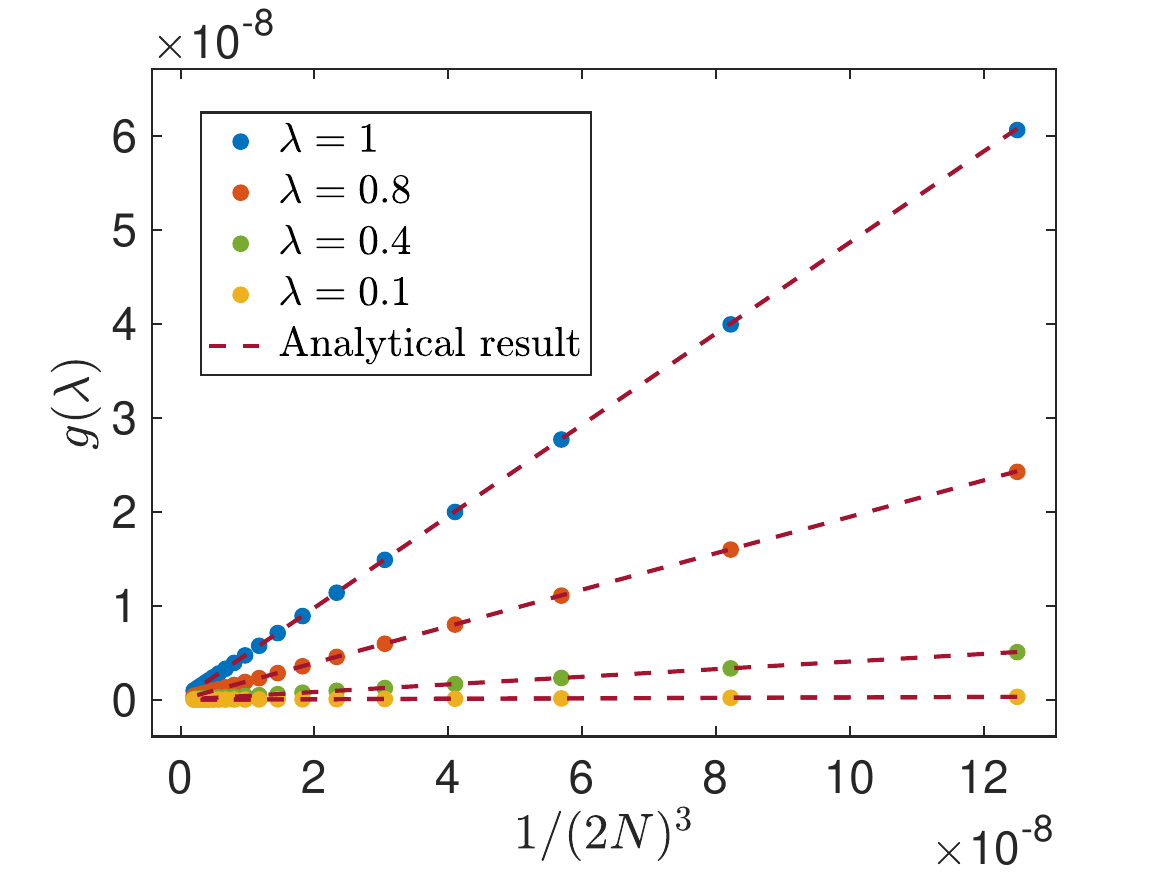}
    \caption{The Liouvillian gap $g(\lambda)$ as a function of the total system size $L=2N$, with local dissipation applied at the left boundary ($n_d = 1$). For various values of the interface parameter $\lambda$, there is an excellent agreement between the numerical results (dots) and the analytical result from perturbative analysis (dashed line) in Eq.~\eqref{eq:perturb_gap_boundary}. Here the total system size is chosen as $L=2N = 200,\, 260, \,320, \dots, 800$, and the dissipation strength is set to be $\gamma = 0.05$.}
    \label{fig: numerical result site 1}
\end{figure}

The Liouvillian gap $g(\lambda)$ is in general a complicated function of the (quasi-)momentum $k$, the dissipative site $n_d$, the total system size $2N$ and the interface parameter $\lambda$. 
Nevertheless, for a given $n_d$, we can always obtain a closed form for $g(\lambda)$. Let us first consider the case of boundary dissipation, i.e. $n_d = 1$. 
Under the limit $N \gg 1$, we have the following asymptotic expression of the Liouvillian gap
\be \label{eq:perturb_gap_boundary}
\begin{aligned} 
g_{n_d=1}(\lambda) & \approx \gamma \min \left\{ 
\frac{\alpha_k^2 k^2 \pi ^2}{(2N+1)^3}
\right\} 
= \frac{ \alpha_1^2 \pi^2 \gamma}{(2N+1)^3} 
\\ & 
= (1 - \sqrt{1 - \lambda^2}) \frac{\pi^2 \gamma}{(2N+1)^3} 
+ \mathcal{O}(\gamma^2). 
\end{aligned}
\ee
For $\lambda=1$, i.e., when there is no conformal interface/defect, 
this is consistent with the previously reported universal scaling $g \propto N^{-3}$ in integrable models~\cite{Prosen_2008, znidaric2011_transpot_xxx, Kehrein2014_cubic_diss, znidaric2015_dissipation_relax, Shibata2020, Yamanaka2021, Tarantelli2021, chenshu2023}. 
What is remarkable here is that the same scaling behavior 
holds even when there is a conformal interface!

In Figure~\ref{fig: numerical result site 1}, we calculate the Liouvillian gap numerically by diagonalizing the structure matrix $\textbf{A}$ defined in Eq.~\eqref{eq:structure_matrix_A}. For various choices of the interface parameter $\lambda$, the numerical result agrees with the analytical solution in Eq.~\eqref{eq:perturb_gap_boundary} very well.

\begin{figure}
    \centering
    \includegraphics[width=0.8\linewidth]{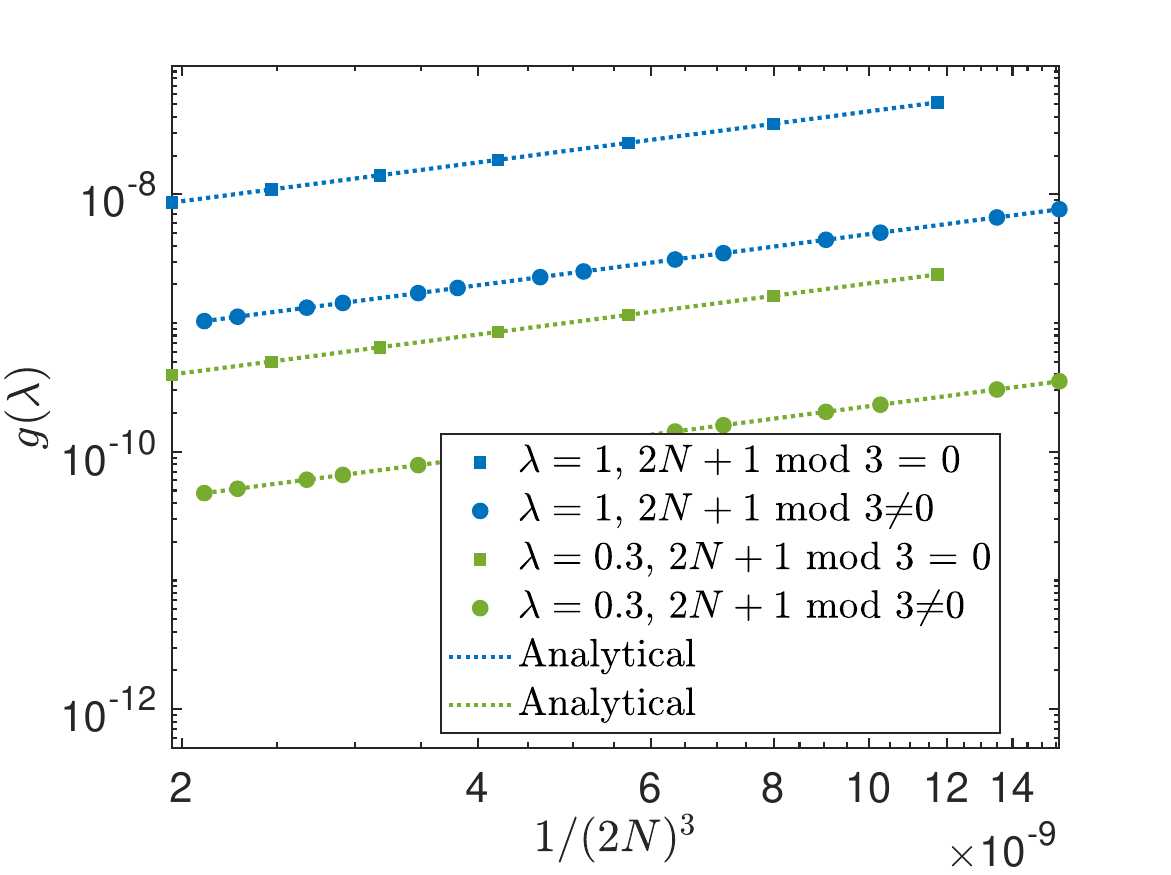}
    \caption{The Liouvillian gap $g(\lambda)$ as a function of the total system size $L=2N$, with local dissipation applied at $n_d = N-1$, the site right next to the interface. For various values of the interface parameter $\lambda$, we find an excellent agreement between the numerical results (dots) and the analytical results (dashed line) in Eq.~\eqref{eq:gapcloseinterface}. Different from the boundary dissipation case, the result here depends on whether $2N+1 \bmod 3 = 0$. Here the total system size is chosen as $2N = 400, 420, 440, \dots, 800$, and the dissipation strength is chosen as $\gamma = 0.05$. }
    \label{fig: ana middle}
\end{figure}

The second case we consider is $n_d=N-1$, i.e., the dissipation is 
applied to the site right next to the interface.
Since the sine function in Eq.~\eqref{eq:Liouvillian_gap_perturbation} can no longer be reduced to a simple power function by taking the limit $N \gg 1$, the analytical solution for the Liouvillian gap becomes more subtle and depends on the specific value of the system size, $2N$. Specifically, by substituting $n_d = N-1$ into Eq.~\eqref{eq:Liouvillian_gap_perturbation}, we obtain the following expression for the Liouvillian gap for a general $N$:
\be
g_{n_d=N-1}(\lambda) \approx \frac{\gamma (1 - \sqrt{1-\lambda^2})}{2N+1} \cos^2 \left(\frac{\pi \eta_{N} }{ 2(2N+1) } \right),
\ee
where
\be
\eta_N = \begin{cases} 
2N - 5, & \quad 2N+1 \bmod 3 = 0, \\ 
2N - 1, & \quad \text{otherwise}.
\end{cases}
\ee
Under the limit $N \gg 1$, it gives 
\be \label{eq:gapcloseinterface}
g_{n_d=N-1}(\lambda) \approx (1 - \sqrt{1 - \lambda^2}) \frac{9\pi^2  s^2 \,\gamma}{4(2N+1)^3} ,
\ee
where 
\be
s = \begin{cases}
2, & \quad  2N+1 \bmod 3 = 0, \\ 
\frac{2}{3}, & \quad \text{otherwise}.
\end{cases}
\ee
Similar to the boundary dissipation case, for $n_d=N-1$, the Liouvillian gap also exhibits an asymptotic scaling behavior of $g \propto N^{-3}$, no matter there is a conformal interface or not. 
We compare this analytical solution with the numerical calculations in Fig.~\ref{fig: ana middle}, and find an excellent agreement 
\footnote{ 
As a remark, we want to point out that the case of $2N + 1 \bmod 3 = 0$ differs from the case of $2N+1 \bmod 3 \neq 0$ in hosting oscillatory coherence, i.e. time-dependent modes that never decay.
Here, we are only interested in the relaxation rate of the decay modes.
}.

Moreover, one can show that the cubic scaling $g(\lambda)\propto N^{-3}$ is universal for a general choice of the dissipative site $n_d$. See Appendix~\ref{app:gap_general_loc} for more details. 

\section{Liouvillian gap for boundary dissipation with finite dissipation strength}
\label{sec:finite}

In the previous section, working in the perturbative regime of the dissipation strength $\gamma$, we derived the relation between the Liouvillian gap and the interface parameter $\lambda$ for an arbitrary location of the additional local dissipation. Here, we show that when the dissipation is introduced exactly at the boundary, the asymptotic behavior of the Liouvillian gap at large system sizes can be obtained at finite $\gamma$, rendering the conclusion non-perturbative.

Our derivation follows the approach in Ref.~\cite{Shibata2020, Yamanaka2021}. We consider the eigenvalue problem  $\textbf{H}_{\text{eff}}\boldsymbol{\psi} = E\boldsymbol{\psi}$ which satisfy following conditions. The effective Hamiltonian $\textbf{H}_{\text{eff}}$ has the same expression as \eqref{eq:Heff_nonherm}.
(i) In the bulk, the eigenvalue equation is given by
\be
J\psi_{i-1} + J\psi_{i+1} = E\psi_{i}. \\
\ee
(ii) At the two boundaries, the eigenvalue problem satisfies
\begin{equation}
\left\{
\begin{split}
 -i\dfrac{\gamma}{2}\psi_{1} + J\psi_{2} =& E\psi_{1} \\
 J\psi_{2N-1} =& E\psi_{2N}
 \end{split}
 \right.
\end{equation}
Thus, one can introduce plane wave ansatz similar to \cite{Shibata2020}
\be
\psi_{j} = Az^{j} + Bz^{-j},
\ee
where $z\in \mathbb{C}.$, and the eigenvalue can be obtained as 
\be
E = -\frac{1}{2}(z + z^{-1}).
\ee

However, in this work we also have a defect at the middle of the system thus it modifies the quantization. (iii) Eigenvalue problem should also satisfy additional conditions due to the defect at the middle of the chain
\begin{align}
   - \frac{1}{2}\psi_{N-1} + \dfrac{\sqrt{1-\lambda^2}}{2}\psi_{N} - \dfrac{\lambda}{2}\psi_{N+1} = E\psi_{N}, \\
    -\dfrac{\lambda}{2}\psi_{N} - \dfrac{\sqrt{1-\lambda^2}}{2}\psi_{N+1} - \frac{1}{2}\psi_{N+2} = E\psi_{N+1}.
\end{align}
So, in this work we introduce the following plane wave ansatz 
\begin{align}
\label{eq:ansatz}
    \psi_{j}(k) =
   \begin{cases}
       \alpha_{k}(Az^{j} + Bz^{-j}) \qquad\quad 1\leq j \leq N, \\
       \beta_{k}(Cz^{j} + Dz^{-j}) \quad N+1\leq j \leq 2N, 
   \end{cases} 
\end{align}
where $\alpha_{k}$ and $\beta_{k}$ has the same form as \eqref{eq:alpha}  and we choose $z=\text{exp}({i\theta_{k}})$ with $k = 1,2,...2N$.\footnote{$\theta_{k}$ can have complex values.} From boundary and defect conditions we obtain the following equations
\begin{equation}
\begin{split}
     &A(1-i\gamma z) + B(1-i\gamma z^{-1}) = 0,\\
     &A(z^{N+1}\alpha_{k} +z^{N}\alpha_{k}\delta) + B(z^{-N-1}\alpha_{k} +z^{-N}\alpha_{k}\delta)  + \\    & C(-z^{N+1}\beta_{k}\lambda) +D(-z^{-N-1}\beta_{k}\lambda) = 0, \\
     &A(-z^{N}\alpha_{k}\lambda) + B(-z^{-N}\alpha_{k}\lambda) +C(z^{N}\beta_{k} -z^{N+1}\beta_{k}\delta) +\\  
     &D(z^{-N}\beta_{k} -z^{-L-1}\beta_{k}\delta) = 0, \\
     &C(z^{2N+1}) + D(z^{-2N-1}) = 0. 
     \end{split}
\end{equation}
where $\delta = \sqrt{1-\lambda^2}$. The above equations have non-trivial solutions when  $\theta_{k}$ satisfy the following quantization condition 
\begin{equation}
    \text{sin}(2N+1)\theta_{k} -i\gamma\text{sin}(2N)\theta_{k} + i\gamma\sqrt{1-\lambda^2}\text{sin}\theta_{k} = 0.
\label{eq:quant}
\end{equation}

We know that for the boundary dissipation Liouvillian gap corresponds to $k=1$ case.
Now, similar to previous works \cite{Shibata2020, Yamanaka2021}, we set $\theta_{1} = \dfrac{\pi}{(2N+1)} + \dfrac{\theta_R+i\theta_I}{(2N+1)^{2}}$. After substituting $\theta_{1}$ in \eqref{eq:quant} and keeping terms upto leading order terms in $1/N$, we get 
\begin{align}
    \theta_{R} = \dfrac{\pi\gamma^2(1-\sqrt{1-\lambda^2})}{1+\gamma^{2}}, \\
    \theta_{I} = -\dfrac{\pi\gamma(1-\sqrt{1-\lambda^2})}{1+\gamma^{2}}.
\label{eq:theta}
\end{align}
Now, we can calculate the eigenvalue 
\begin{equation}
\begin{split}
    E =& -\frac{1}{2}(z+z^{-1})  \\
    =& -\dfrac{1}{2}\text{exp}\left(\dfrac{i\pi}{2N+1}\right)\left(1-\dfrac{\theta_{I}}{2N+1}\right) \\
    & -\dfrac{1}{2}\text{exp}\left(\dfrac{-i\pi}{2N+1}\right)\left(1+\dfrac{\theta_{I}}{2N+1}\right) \\
    = &-\text{cos}\left(\dfrac{\pi}{2N+1}\right) +\dfrac{i\theta_{I}}{(2N+1)}\text{sin}\left(\dfrac{\pi}{2N+1}\right).
\end{split}
\end{equation}
Substituting $\theta_{I}$ from \eqref{eq:theta}, we get the maximum imaginary part of eigenvalues as\footnote{For the Liouvillian gap $g = - 2 \max_{\Im[E_{i}] \neq 0} \left\{\Im[E_{i}] \right\}$, where $E_i$ are the eigenvalues of $\textbf{Z}$. The effective Hamiltonian matrix is related to the matrix $\textbf{Z}$ as $\textbf{Z} = \frac{1}{2} \textbf{H}_{\text{eff}}$ (See \eqref{eq:Heff_nonherm}). Hence, the Liouvillian gap is  
$g = -\text{max}\{E_{\text{imag}}\}$.}
\begin{equation}
    \text{max}\{E_{\text{imag}}\} = - \dfrac{\pi\gamma(1-\sqrt{1-\lambda^2})}{(1+\gamma^{2})(2N+1)^{2}}\text{sin}\left(\dfrac{\pi}{2N+1}\right).
\end{equation}
Thus, for $2N>>1$, we find the expression of Liouvillian gap for finite values of dissipation strength $\gamma$ as
\begin{align}
    g(\lambda) = - \text{max}\{E_{\text{imag}}\} 
     = \dfrac{\gamma\pi^{2}(1-\sqrt{1-\lambda^2})}{(1+\gamma^{2})(2N+1)^{3}}.
\end{align}
Finally, we note that for local dissipation away from the boundary, the plane-wave ansatz in Eq.~\eqref{eq:ansatz} must be modified, which in turn alters the quantization condition in Eq.~\eqref{eq:quant}. In this case, the Liouvillian gap does not necessarily correspond to the $k=1$ mode.

\section{Suppression of relaxation for characterizing conformal interface} 
\label{sec:suppress}

Based on the discussion of Liouvillian spectrum and Liouvillian gap, in this section, we are ready to study how a conformal interface suppresses the relaxation rate in a dissipative system.

As we have discussed in the introductory part, in a closed critical system, there are two typical quantities -- $c_{\text{LR}}$ and $c_{\text{eff}}$ --  that characterize the transmission properties of a conformal interface. They measure how the energy transmission and the information transmission are suppressed across a conformal interface, respectively. In an open system, however, neither energy nor information is conserved, and the quantum dynamics becomes non-unitary. To characterize the conformal interface in this case,  we consider the Liouvillian gap $g$, a fundamental property in dissipative quantum systems.

In the previous section, we have shown that the Liouvillian gap exhibits a universal scaling $g(\lambda) \propto 1 / N^3$ in a critical free fermionic chain with conformal interface. Although the concrete form of 
$g(\lambda)$ depends on the dissipation strength $\gamma$, the dissipation location $n_d$, and the total length $L=2N$ in a complicated way [see, e.g., \eqref{eq:perturb_gap_boundary} and \eqref{eq:gapcloseinterface}], what is remarkable, the ratio of the Liouvillian gaps with and without interface is universal as follows
\be
\label{g_ratio}
\frac{g(\lambda)}{g(\lambda=1)} \approx 1 - \sqrt{1 - \lambda^2} , 
\ee
which is calculated to first order in perturbation theory  in $\gamma$ (and also for the finite value of $\gamma$ when the dissipation is located at the boundary of a long critical chain). That is, \eqref{g_ratio} only depends on the interface parameter $\lambda$.
This motives us to define the relaxation coefficient $c_{\text{relax}}$ as
\be
\frac{c_{\text{relax}}}{c} := \dfrac{g(\lambda)}{g(\lambda=1)} , 
\ee
where $c$ is the central charge of the defect-free critical system at $\lambda=1$. In the free fermion model we study, we have $c = 1$, leading to 
\be \label{eq:crelax_analytic}
c_{\text{relax}} \approx 1 - \sqrt{1 - \lambda^2} =: \, c_{\text{relax}}^{(1)}  , 
\ee
where $c_{\text{relax}}^{(1)}$ denotes the approximation of $c_{\text{relax}}$ up to the first order perturbation. It is expected that \eqref{eq:crelax_analytic} holds for 
 small dissipation strength $\gamma$  in the perturbative regime $\gamma\ll J$, where $J=1$ is the hopping strength in \eqref{eq:H1}. Moreover, when local dissipation is introduced precisely at one of the boundaries, we expect the result \eqref{eq:crelax_analytic} to hold for any finite dissipation strength $\gamma$.
Physically, $c_{\text{relax}}$ captures how the relaxation rate in dissipative systems is suppressed by the presence of a conformal interface, similar to how the energy and information transmissions are suppressed in closed systems. 
As shown in Fig.~\ref{fig:MainResult}, we compare $c_{\text{relax}}$ with $c_{\text{LR}}$ and $c_{\text{eff}}$ in the critical free fermion chain. 
One can find they show distinct features from each other -- this indicates that $c_{\text{relax}}$ is indeed a 
new quantity characterizing the conformal interface in a dissipative system. 

\medskip
We also compare the analytical result in \eqref{eq:crelax_analytic} with the numerical results from exact diagonalization of the structure matrix $\textbf{A}$. As shown in Fig.~\ref{fig:MainResult}, they agree in a remarkable way for various choices of dissipation locations $n_d$. 
This suggests that $c_{\text{relax}}$ well captures universal property of the conformal interface in our dissipative lattice model.

\begin{figure}
    \centering
    \begin{tikzpicture}
            \node[inner sep=0pt] (russell) at (-50pt, -85pt)
    {\includegraphics[width=.27\textwidth]{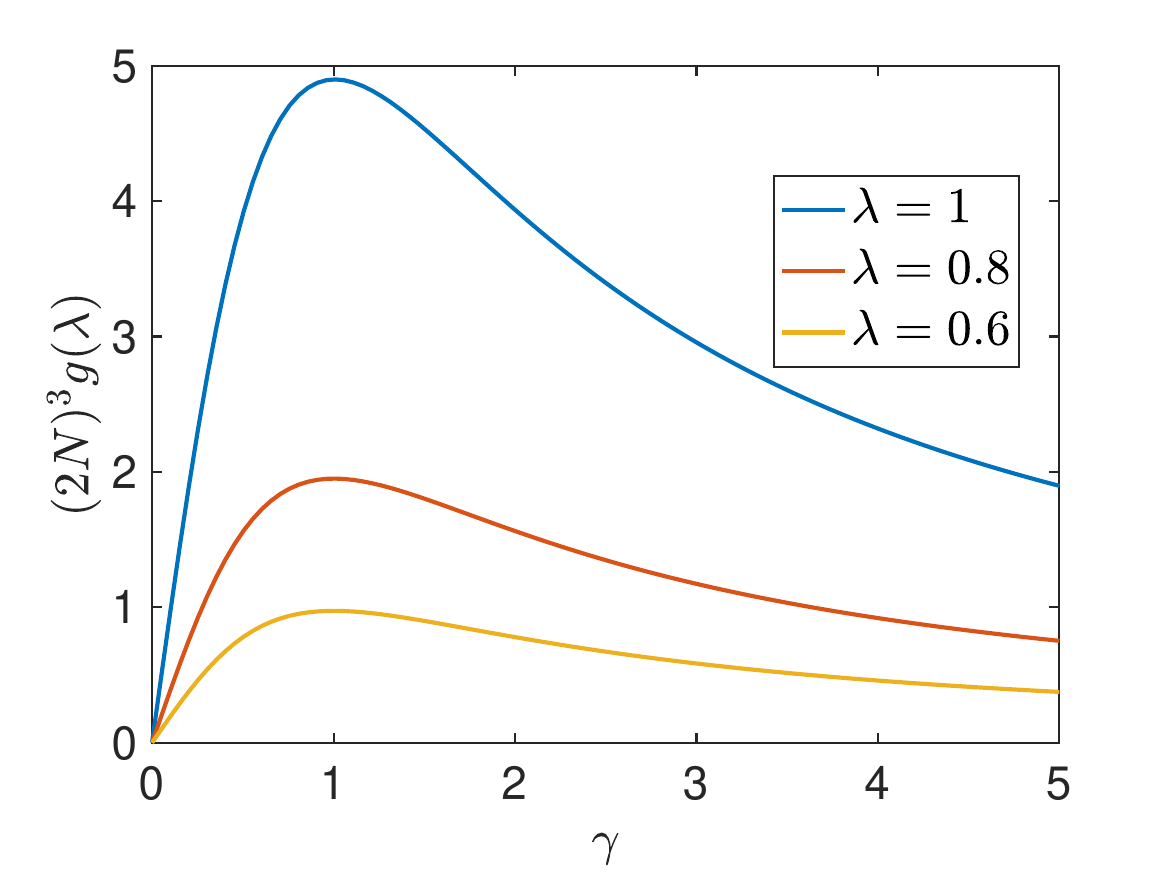}};
        \node[inner sep=0pt] (russell) at (82pt,-85pt)
    {\includegraphics[width=.27\textwidth]{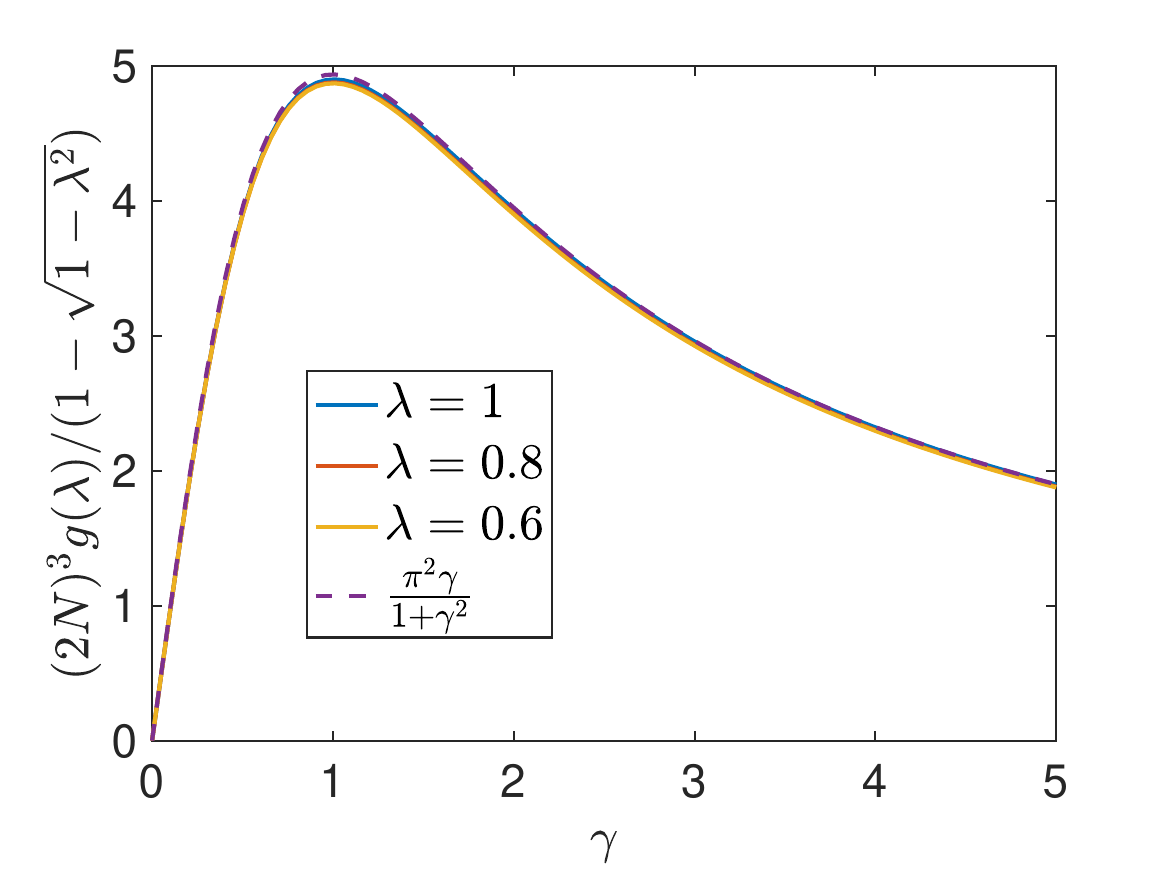}};
            \node[inner sep=0pt] (russell) at (-50pt, -187pt)
    {\includegraphics[width=.27\textwidth]{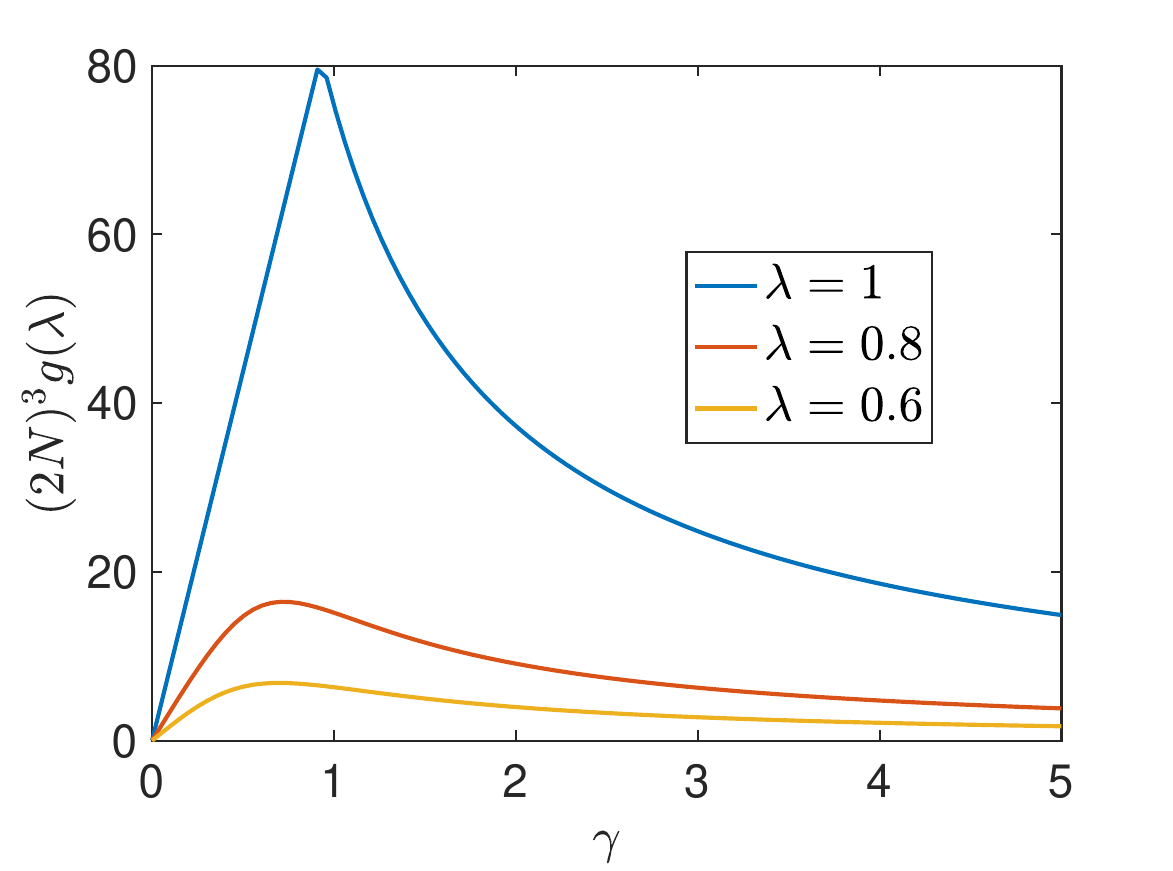}};
        \node[inner sep=0pt] (russell) at (80pt,-187pt) 
    {\includegraphics[width=.27\textwidth]{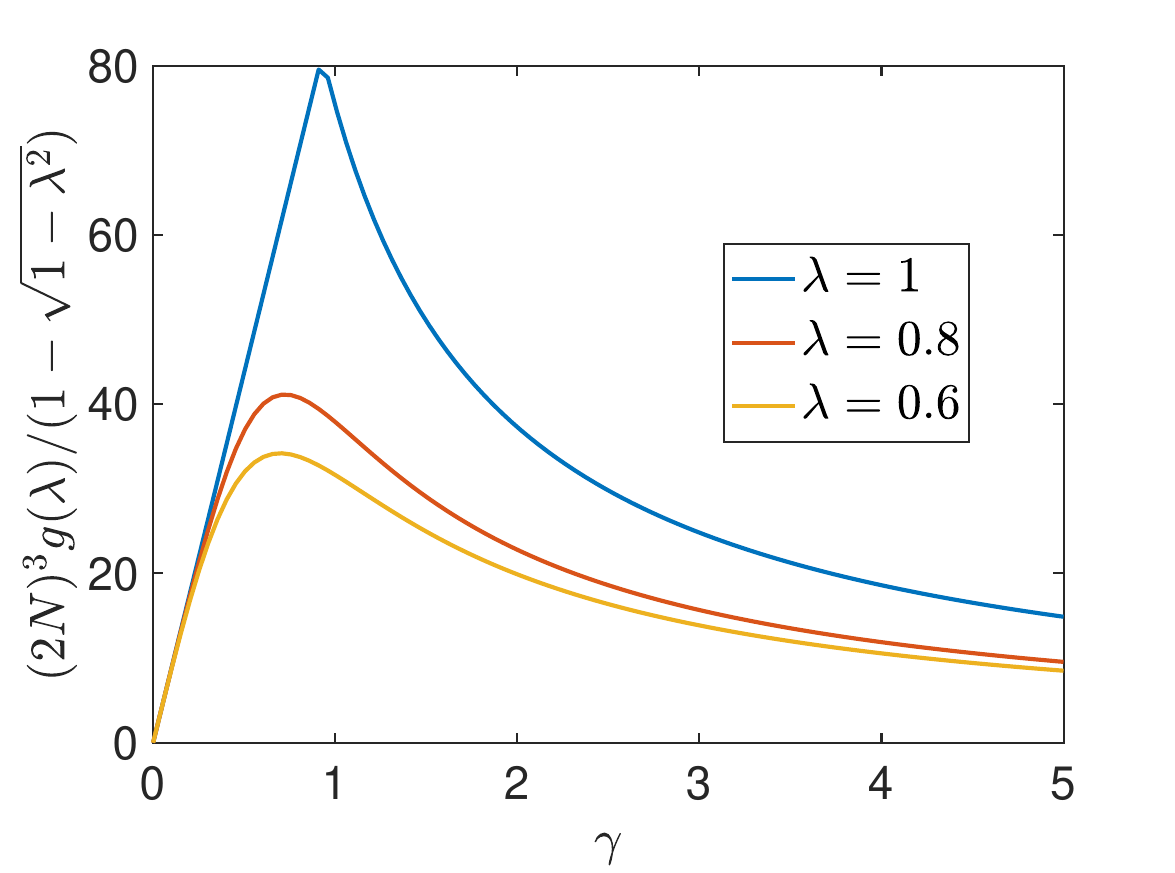}};
            \node at (-45pt, -55pt){(a)};
            \node at (120pt,-55pt){(b)};
            \node at (-20pt, -155pt){(c)};
            \node at (120pt,-155pt){(d)};
    \draw[red] (1.02,-7.0) rectangle (1.42,-7.8) ;       
\end{tikzpicture}
\caption{Numerical results for the rescaled Liouvillian gap $(2N)^{3}g(\lambda)$ and $(2N)^{3}g(\lambda)/c^{(1)}_{\text{relax}} = (2N)^{3}g(\lambda)/(1 - \sqrt{1 - \lambda^2})$, plotted as a function of the dissipation strength $\gamma$. (a) and (b): The local dissipation is applied at the boundary, i.e. at $n_d = 1$. (c) and (d): The local dissipation is applied right next to the interface, i.e. at $n_d = N-1$. In (b), the rescaled Liouvillian gap collapses to a single curve of $\pi^2 \gamma / (1+\gamma^2)$, denoted by the dashed line. In (d), the weak-dissipation regime is highlighted, where $c_{\text{relax}}$ exhibits a linear dependence on $\gamma$ and all curves collapse to the same one in this regime. The total system size is chosen as $2N = 200$.} 
\label{fig:scaling}
\end{figure}
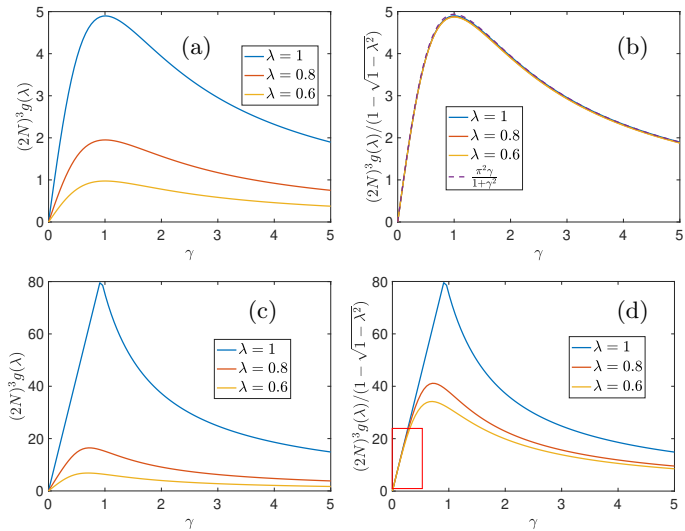

Furthermore, as shown in Fig.~\ref{fig:scaling}, we numerically calculate the Liouvillian gap for finite values of the dissipation strength $\gamma$. This allows us to study the Liouvillian gap out of the weak dissipation regime.
Interestingly, we observe a universal cubic scaling $g \propto 1/N^3$ for arbitrary values of $\gamma$. In Fig.~\ref{fig:scaling}, we present the numerical results for the rescaled Liouvillian gap $N^3 g(\lambda)$ and  its ratio to the relaxation coefficient $c_{\text{eff}}^{(1)} = 1 - \sqrt{1 - \lambda^2}$, as a function of the dissipation strength $\gamma$ for different choices of the interface parameter $\lambda$.
One important feature we observe is that for the boundary dissipation case with $n_d=1$, 
all curves of $(2N)^{3}g(\lambda)/c^{(1)}_{\text{relax}}$
collapse to a single curve described by $\pi^2 \gamma / (1+\gamma^2)$ \cite{Shibata2020}. This means our result of $c_{\text{relax}}$ in \eqref{eq:crelax_analytic} holds for finite dissipation strengths $\gamma$ that are out of the perturbation regime.


As we move the dissipation location $n_d$ from the boundary into the bulk, however, we find that the collapse of curves only happens in the weak dissipation regime where $\gamma\ll J$, as shown in Fig.\ref{fig:scaling} (d). This means for a general choice of dissipation location $n_d$, our result in 
\eqref{eq:crelax_analytic} only holds in the weak dissipation regime.

\medskip
As a short summary, by comparing with the exact numerical calculation, we have verified our analytical result in 
\eqref{eq:crelax_analytic} in the weak dissipation regime. 
That is, $c_{\text{relax}}$ is 
independent of both the dissipation strength $\gamma$ and the 
dissipation location $n_d$, as long as the dissipation is weak ($\gamma\ll J$). Very interestingly, when the dissipation is at the boundary, our result in \eqref{eq:crelax_analytic} holds for arbitrarily finite dissipation strengths $\gamma$.

\section{Discussion and conclusion} \label{sec:discuss}

In this work, we investigate the effect of conformal interface on the relaxation dynamics of a dissipative quantum critical system.
By studying a free fermion lattice at the critical point with a local dissipation, we find that one needs to introduce a new quantity $c_{\text{relax}}$, which we call relaxation coefficient, to characterize the conformal interface in an open system. Physically, $c_{\text{relax}}$ measures how the relaxation rate is suppressed by the conformal interface. 
Our result for $c_{\text{relax}}$ is universal in the sense that it is independent of the dissipation strength in the weak dissipation limit, and it is also independent of the location of the dissipation. Interestingly,
when the local dissipation is at the boundary, our analytical result for 
$c_{\text{relax}}$ remains valid even for finite dissipation strengths that 
are out of the weak dissipation regime.

\medskip
We find that $c_{\text{relax}}$ shows a distinct behavior from the previously known quantities $c_{\text{eff}}$ and $c_{\text{LR}}$, which characterize the transmission properties of conformal interface in a closed system. We propose the inequalities for $c_{\text{relax}}$, $c_{\text{eff}}$ and $c_{\text{LR}}$ in \eqref{eq:MainResult}, which 
are interesting to test in general CFTs in the future.
Our work not only provides a solvable example of 
how a conformal interface suppresses the relaxation rate in an open quantum critical system, but also offers potential insights toward establishing a general theory for characterizing conformal interfaces subjected to dissipation.

\bigskip

There are many interesting future directions related to our setup in this work, and we mention a few of them here:

-- \textit{Conformal interfaces in general CFTs with local dissipation:}
In this work, we focus on the effect of conformal interface in a dissipative free-fermion chain at the critical point. 
There are other interesting critical lattice models where the conformal interfaces are well studied, when the system is closed
\cite{oshikawa1997_Ising_defect,2012_Peschel_Eisler}. 
It would be interesting to study explicitly the effect of conformal interfaces in these models in the presence of local dissipations, 
and check whether the relation 
$0\le c_{\text{relax}}\le c_{\text{LR}}\le c_{\text{eff}}$
found in this work still holds in other lattice models.
In addition, there are more general cases where the CFTs on the two 
sides of the conformal interface are different.
For instance, one can consider the conformal interface between free bosonic CFTs with different compactification radius~\cite{Kazuhiro_Sakai_2008, Roy2023_interface_LL} (different CFTs with the same central charge $c=1$) and the conformal interface between nearest minimal models~\cite{Quella2006_transmission, Gaiotto2012_RG_DW, Tang2023, Cogburn2023} (different CFTs with distinct central charges). These models offer useful platforms for studying how conformal interfaces influence the dissipative dynamics of broader classes of critical theories, a direction we leave for future exploration.

\medskip
-- \textit{String order parameter to detect boundary symmetry breaking in open systems:} Recently, it was found that for a closed quantum critical system, where the global symmetry is explicitly broken only at the boundary or interface, one can use string order parameters to detect such boundary symmetry breaking \cite{2025_Barad}. In particular, the time evolution of the string order parameter shows universal features that are independent of the concrete lattice models. In our setup, the boundary dissipation breaks the strong $U(1)$ symmetry explicitly. It is interesting to study whether the string order parameter proposed in closed systems can be generalized to this open system and whether there are universal features in the time evolution. In addition, since a conformal interface will suppress the relaxation rate toward the steady state, it is expected that it will affect the time evolution of string order parameters.

\medskip
-- \textit{Holographic interface CFTs with dissipation:}
In recent years, the study of interface CFTs
from the AdS/CFT correspondence has
 provided many insights in our understanding of 
 general properties of conformal interfaces
\cite{
Myers2020_defect_island_1, Myers2020_defect_island_2, Sonner2022_Island,
2008_Takayanagi,Karch2024,Tang2023,2021_Luo,2023_Karch,2023_Karch_effectiveC,2024_Basak}.
For example, the remarkable relation in \eqref{Eq:bound} was verified in holographic interface CFTs based on the AdS/CFT correspondence \cite{Karch2024}. So far, related studies are mainly focused on closed systems. 
In the dissipative critical fermionic model as studied in this work, we observe that the new quantity $c_{\text{relax}}$ satisfies the relation in \eqref{eq:MainResult} (see also Fig.~\ref{fig:MainResult}). It would be interesting to investigate the property of $c_{\text{relax}}$ from the AdS/CFT correspondence point of view and study whether the relation 
in \eqref{eq:MainResult} holds in general \footnote{We thank Andreas Karch for an interesting discussion on this topic.}.
In a very recent work \cite{2025_Ishii}, the authors studied the AdS/CFT correspondence for the Lindbladian dynamics of a dissipative CFT. The method therein may be applied to the study of holographic interface CFTs.

\acknowledgments

We thank for the interesting discussions with Po-Yao Chang, Andreas Karch, Zhu-Xi Luo, and Carolyn Zhang, and we thank Andreas Karch for letting us know the reference \cite{2020_Meineri}.
This work is supported by a startup at Georgia Institute of Technology.

\appendix


\section{ Lattice calculation of $c_{\text{LR}}$}
\label{Appendix:c_LR}

In this appendix, we show how to extract the
transmission coefficient $c_{\text{LR}}$ from a lattice model based on the ``scattering experiment'' proposed in Ref.~\cite{2020_Meineri}. With this method, one can see clearly why $c_{\text{LR}}$ measures the \textit{suppression} of energy transmission by the conformal interface.

The method proposed in Ref.~\cite{2020_Meineri} can be sketched as follows: 
\be
\label{CFT_collider}
\begin{tikzpicture}[x=0.75pt,y=0.75pt,yscale=-0.8,xscale=0.8]

\small
\draw    (141,165) -- (191,115.33) ;
\draw [shift={(168.7,137.49)}, rotate = 135.19] [fill={rgb, 255:red, 0; green, 0; blue, 0 }  ][line width=0.08]  [draw opacity=0] (7.14,-3.43) -- (0,0) -- (7.14,3.43) -- cycle    ;
\draw    (191,115.33) -- (142.33,65.33) ;
\draw [shift={(164.02,87.61)}, rotate = 45.77] [fill={rgb, 255:red, 0; green, 0; blue, 0 }  ][line width=0.08]  [draw opacity=0] (7.14,-3.43) -- (0,0) -- (7.14,3.43) -- cycle    ;
\draw    (191,115.33) -- (241,65.67) ;
\draw [shift={(218.7,87.82)}, rotate = 135.19] [fill={rgb, 255:red, 0; green, 0; blue, 0 }  ][line width=0.08]  [draw opacity=0] (7.14,-3.43) -- (0,0) -- (7.14,3.43) -- cycle    ;
\draw [color={rgb, 255:red, 155; green, 155; blue, 155 }  ,draw opacity=1 ][line width=2.25]    (191,46.67) -- (191,180.67) ;
\draw    (227,142.67) -- (227,170) ;
\draw [shift={(227,140.67)}, rotate = 90] [fill={rgb, 255:red, 0; green, 0; blue, 0 }  ][line width=0.08]  [draw opacity=0] (12,-3) -- (0,0) -- (12,3) -- cycle    ;
\draw    (255.67,169.38) -- (227,170) ;
\draw [shift={(257.67,169.33)}, rotate = 178.75] [fill={rgb, 255:red, 0; green, 0; blue, 0 }  ][line width=0.08]  [draw opacity=0] (12,-3) -- (0,0) -- (12,3) -- cycle    ;

\draw (140.67,45.07) node [anchor=north west][inner sep=0.75pt]    {$\text{CFT}_L$};
\draw (206,45.07) node [anchor=north west][inner sep=0.75pt]    {$\text{CFT}_R$};
\draw (130,100) node  {reflected};
\draw (255,100) node {transmitted};
\draw (125.33,147) node {incident};
\draw (234,133.07) node [anchor=north west][inner sep=0.75pt]    {$t$};
\draw (260,158.4) node [anchor=north west][inner sep=0.75pt]    {$x$};

\end{tikzpicture}
\ee
First, we generate an energy pulse, which is right-moving here, and let it collide with the interface. As the energy pulse passes through the conformal interface, it will be
scattered into a transmitted energy pulse and a reflected energy pulse.
The energy transmission/reflection coefficients of the conformal interface can be defined as
\be
\label{Transmission_energy}
\mathcal T=\frac{\text{transmitted energy}}{\text{incident energy}},
\quad
\mathcal R=\frac{\text{reflected energy}}{\text{incident energy}}.
\ee
We have $\mathcal T+\mathcal R=1$ due to energy conservation
and $\mathcal T, \,\mathcal R\ge 0$ due to the positivity of transmitted and reflected energy.
Remarkably, as shown in Ref.~\cite{2020_Meineri}, $\mathcal T$ and $\mathcal R$ are completely independent of the details of the incoming excitation.
With this universal property, one can use \eqref{Transmission_energy}
to define $c_{\text{LR}}$ as follows:
\be
\label{Transmission_cLR}
\mathcal{T}_L=\frac{ c_{\text{LR}} }{ c_{L} }, \quad \mathcal{T}_R=\frac{ c_{\text{LR}} }{ c_{R} }.
\ee
Here the subscript of $\mathcal{T}$ indicates the origin of the excitation, i.e., $L$ represents right-moving excitation (from the left) and $R$ represents left-moving excitation (from the right).

\begin{figure}[tp]
\centering
\begin{tikzpicture}
  \node[inner sep=0pt] (russell) at (0pt,0pt)
    {\includegraphics[width=2.25in]{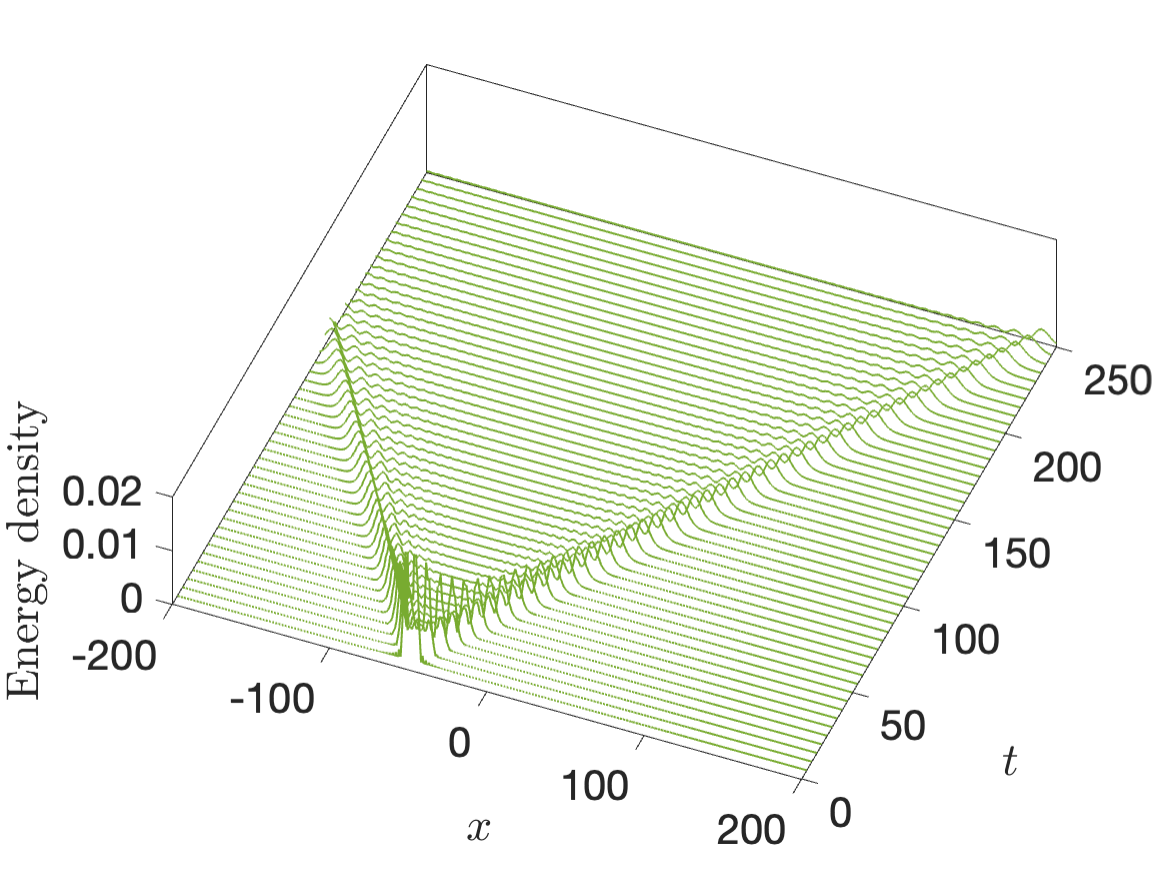}};

     \node[inner sep=0pt] (russell) at (0pt,-120pt)
    {\includegraphics[width=2.27in]{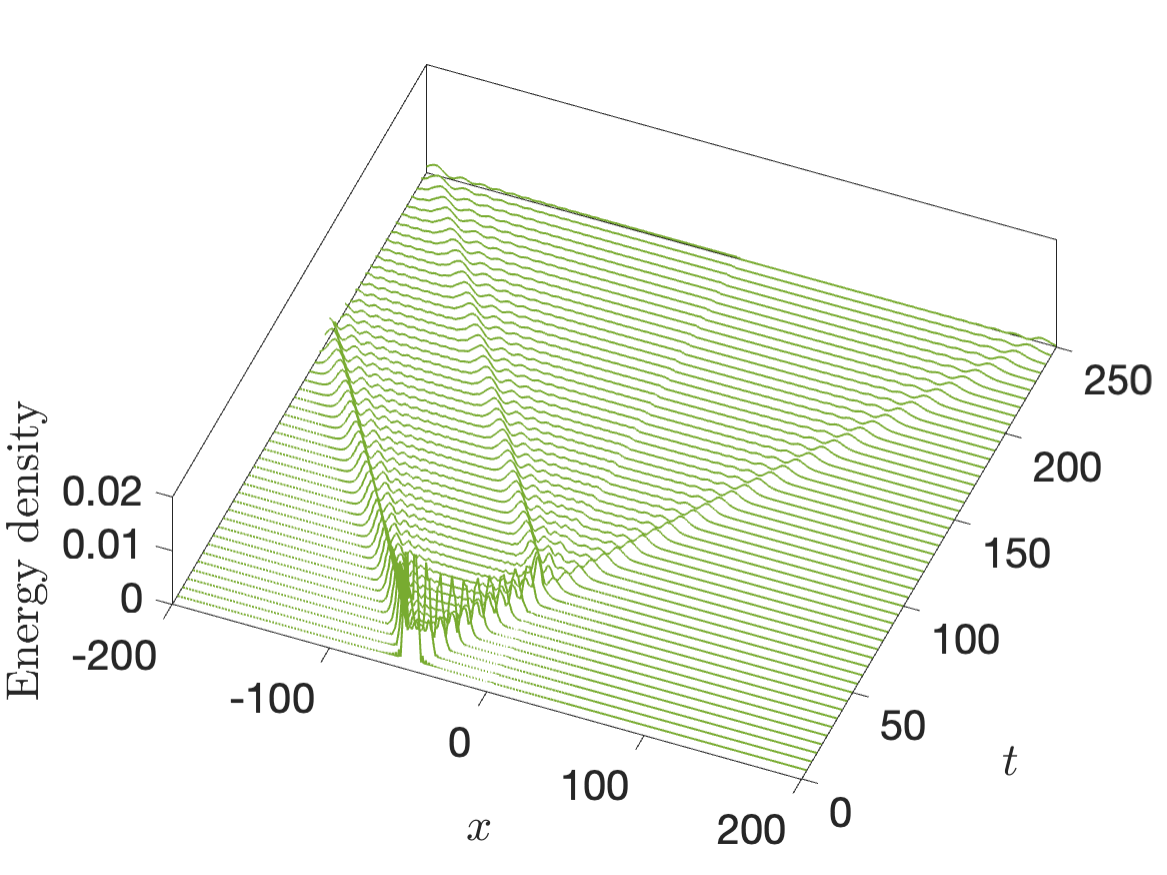}};
  
      \node[inner sep=0pt] (russell) at (0pt,-240pt)
    {\includegraphics[width=2.25in]{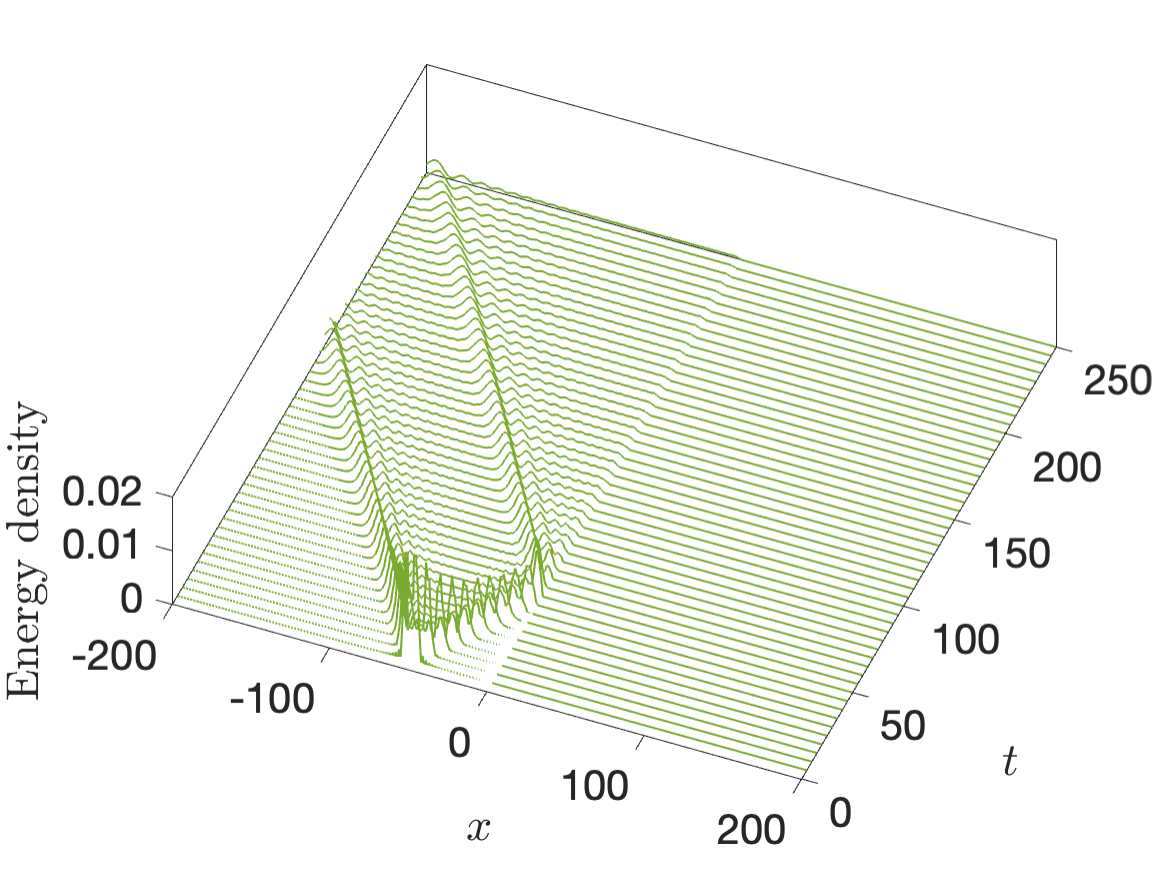}};
    
 \node at (30pt,50pt){\textcolor{black}{Total transmission}};

  \node at (30pt,-75pt){\textcolor{black}{Partial transmission}};

  \node at (35pt,-200pt){\textcolor{black}{Total reflection}};
 
    \end{tikzpicture}
    \caption{Energy scattering by a conformal interface after a local quantum quench.
    The sample plots here correspond to the conformal interfaces with total transmission ($\lambda=1$), partial transmission/partial reflection ($\lambda=0.6$), and total reflection ($\lambda=1$). Here $\lambda$ is introduced in \eqref{eq:H1} and \eqref{eq:H2}.
    We join the two decoupled chains at $x=-50$, and the conformal interface is inserted at $x=0$. The two decoupled chains before joint together are defined on $[-400,-50]$ and $[-50,400]$, respectively. 
    }
 \label{Fig:EnergyScatter}
\end{figure}
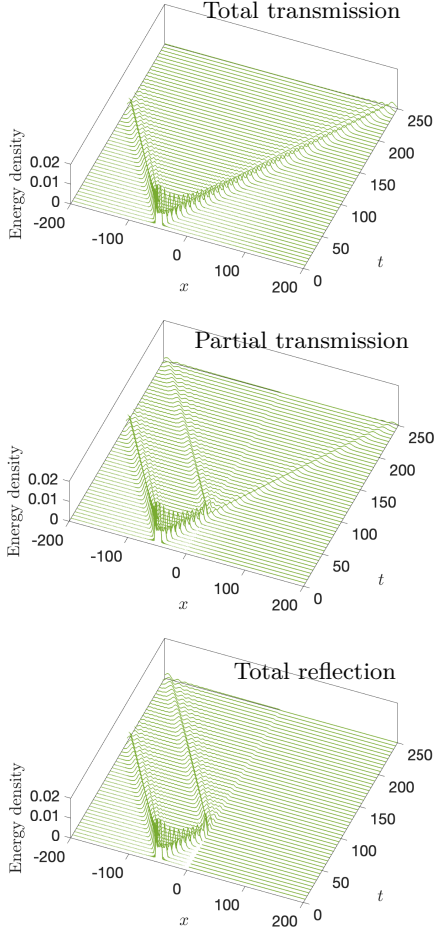

Next, we will use \eqref{Transmission_energy} and \eqref{Transmission_cLR} to extract $c_{\text{LR}}$ on a
lattice model. The incident energy pulse in \eqref{CFT_collider} will be generated by a local quantum quench.
More concretely, we consider two decoupled free-fermion chains, one defined on sites from  $-N$ to $-N+N_1$, and the other defined on sites from $-N+N_1+1$ to $N$.

The conformal interface is inserted locally at two sites $0$ and $1$. That is, the Hamiltonians for the two decoupled chains are $H_1$ and $H_2$, where
\be
\label{H1_appendix}
H_1=-\frac{1}{2}\sum_{i=-N}^{-N+N_1-1} c_i^\dag c_{i+1}+h.c.,
\ee
where $\{c_i^\dag,c_j\}=\delta_{ij}$ and $\{c_i,c_j\}=\{c_i^\dag,c_j^\dag\}=0$.
$H_2$ has the same expression as \eqref{eq:H1}, except that 
now the system is defined on sites from $-N+N_1+1$ to $N$ and the conformal defect is defined on sites $0$ and $1$.

\begin{figure}
    \centering
    \includegraphics[width=0.8 \linewidth]{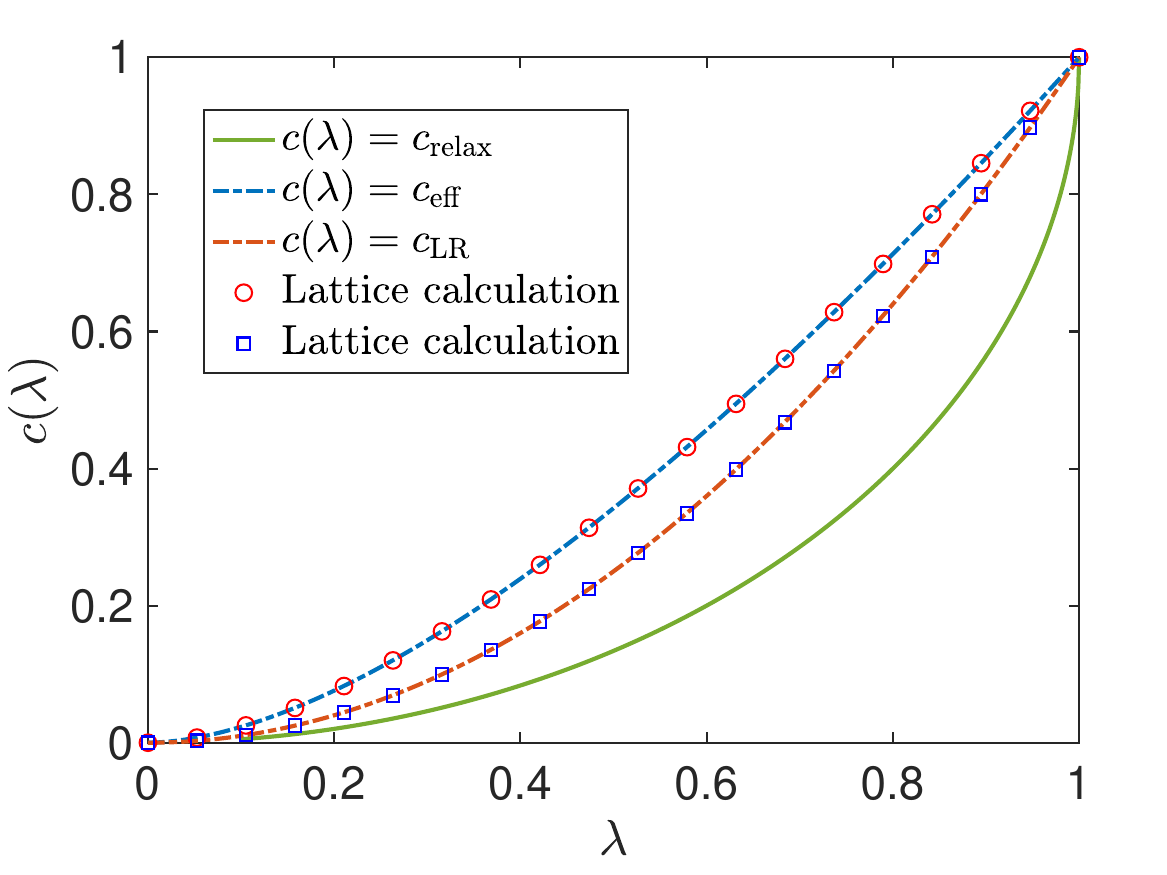}
    \caption{Comparison of $c_{\text{LR}}$ and $c_{\text{eff}}$ obtained from analytical results (dashed lines) in \eqref{eq:cLR_energy} and \eqref{eq:ceff_entangle} and lattice calculations (circles and squares).
    The lattice model calculations of $c_{\text{LR}}$ and $c_{\text{eff}}$ are described in Appendices \ref{Appendix:c_LR} and \ref{Appendix:c_eff}, respectively. Here, as a comparison, the line of $c_{\text{relax}}$ is the perturbative result in Eq.~\eqref{eq:crelax_analytic}. 
    }
    \label{fig:c_LR_lattice}
\end{figure}

Then the initial state $|\psi_0\rangle$ is prepared as the tensor product of the ground states in each chain as $|\psi_0\rangle=|G_1\rangle\otimes |G_2\rangle$.
At $t=0$, we join the two decoupled chains by turning on the coupling at their ends with the total Hamiltonian given by
\be
H=H_1+H_2+H_{\text{int}},
\ee
where $H_{\text{int}}=-\frac{1}{2}c_{N_1}^\dag c_{N_1+1}+h.c.$.
As shown in Fig.\ref{Fig:EnergyScatter}, the excitations are generated from $x=-N+N_1$, where we join the two chains. Note that, here $H_{\rm int}$ matches the bulk hopping term on both the left and right chains and does not define an interface. The conformal defect is instead inserted in the right chain $H_2$.
As time evolves, we have a right-moving pulse and a left-moving pulse. When the right-moving energy pulse arrives at the conformal interface (at $x=0$), it will split into two parts:
one is transmitted and the other is reflected. 
The amount of energy transmitted/reflected depends on the details of the conformal interface, which is characterized by $\lambda$ in \eqref{eq:H2}. As shown in Fig.\ref{Fig:EnergyScatter}, when the conformal interface is totally transmissive ($\lambda=1$), the right-moving energy pulse
propagates freely in space. If the conformal interface is partially transmissive ($0<\lambda<1$), the pulse is partially transmitted and partially reflected. If the interface is totally 
reflective ($\lambda=0$), then one can find that the energy pulse is totally reflected, which becomes left-moving.

Numerically, we calculate the transmission coefficient $\mathcal T$ based on \eqref{Transmission_energy}. Then by using the definition in \eqref{Transmission_cLR}, we can obtain $c_{\text{LR}}$, as shown in Fig.\ref{fig:c_LR_lattice}.
One can find the lattice results and analytical results agree with each other very well.
As a remark, in Fig.\ref{Fig:EnergyScatter} we consider the right-moving energy flux to extract $c_{\text{LR}}$. One can certainly consider the left-moving energy flux, which gives the same result.

\section{Lattice calculation of $c_{\text{eff}}$}
\label{Appendix:c_eff}

\begin{figure}[htp]
    \centering
    \includegraphics[width=0.8\linewidth]{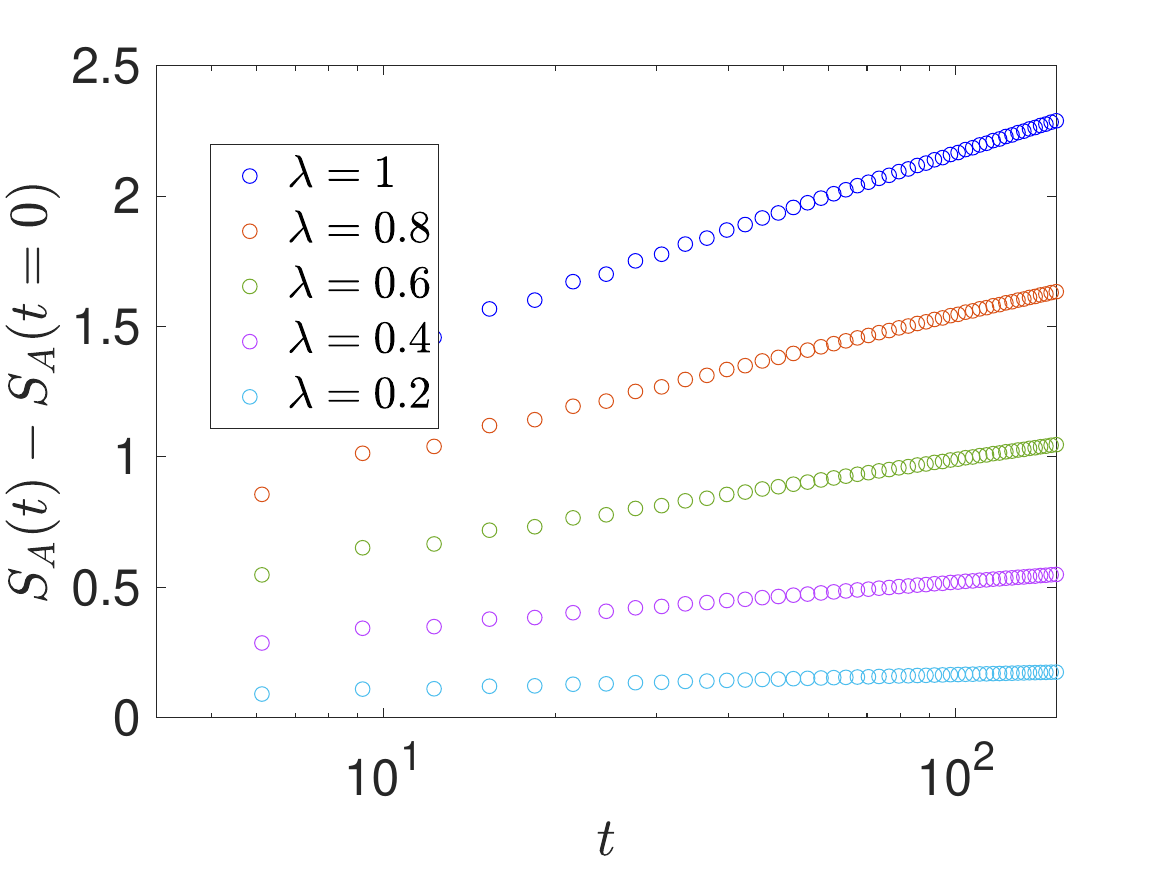}
    \caption{Time evolution of the entanglement entropy for $A=[0,L/2]$ based on a free fermion lattice calculation with different strengths of conformal interfaces. The interface is inserted at $x=L/2$ and the total length of the lattice is chosen as $L=800$.
    }
    \label{fig:c_eff_lattice}
\end{figure}

With a setup similar to that in Appendix \ref{Appendix:c_LR}, one can extract the effective central charge $c_{\text{eff}}$, which characterizes how the quantum entanglement across the conformal interface is suppressed, by studying the time evolution of the entanglement entropy. 

Note that $c_{\text{eff}}$ has been well studied in both CFT calculations and lattice calculations\cite{Kazuhiro_Sakai_2008, Eisler2010, Eisler_2012, brehm_2015_EE_interface_ising, Wen2017, Eisler2022, Tang2023, Karch2023_effective}.
In particular, it was shown in \cite{Wen2017} that the conformal interface suppresses the ground-state entanglement entropy and the time-dependent entanglement entropy after quantum quenches with the same coefficient $c_{\text{eff}}$. Here we follow the method in \cite{Wen2017} and extract $c_{\text{eff}}$ from a lattice free fermion calculation.

First, we consider two decoupled CFTs defined in $[-l,0]$ and $[0,l]$ and join them at their ends at $x=0$ through a conformal interface. As shown in \cite{Wen2017}, the time evolution of entanglement entropy for the subsystem $A=[0,l]$ is described by
\be
S_A(t)-S_A(0)\simeq \frac{c_{\text{eff}}}{3}\log (t).
\ee
To extract $c_{\text{eff}}$, we consider two decoupled free fermion lattices at the critical point. Each Hamiltonian has the form of \eqref{H1_appendix}. Then at $t=0$, we couple the two lattices at their ends with a conformal interface as described in \eqref{eq:H2}.
Then we calculate the time evolution of entanglement entropy for half of the total system, with the entanglement cut chosen right at the interface. 
As shown in Fig.\ref{fig:c_eff_lattice}, one can see clearly a $\log(t)$ growth of entanglement entropy, with the coefficient $c(\lambda)/3$ depending on the strength of the conformal interface. Then we extract the effective central charge $c_{\text{eff}}$ by using $c_{\text{eff}}(\lambda)=c(\lambda)/c(\lambda=1)$, and the result is shown in Fig.\ref{fig:c_LR_lattice}, where one can find an excellent agreement with the CFT calculations.

\section{Liouvillian gap for general locations}\label{app:gap_general_loc}

In the main text, we mainly considered two cases: (1) local dissipation is applied at the boundary ($n_d=1$), and (2) local dissipation applied at the closest site to the interface ($n_d=N-1$). In this appendix, we comment on the case where local dissipation is applied at a general site $n_d$. Recall that based on the third quantization formalism, we have perturbatively derived the following expression of the Liouvillian gap 
\be
\label{eq:gap1}
g(\lambda) \approx \min \left\{\frac{\gamma \alpha^{2}_{k}}{2N+1} \sin^2 \left(\frac{ n_d k\pi}{2N+1} \right) \right\} ,
\ee
where $\alpha_k^2 = 1 + (-1)^k \sqrt{1 - \lambda^2}$, $k = 1, 2, \dots, 2N$.

For general location of the dissipative site $n_d$, it is hard to determine the minimum, since one needs to find the global minimum for both the factor $\alpha_k^2$ (depending on the even-odd parity of $k$) and the sine function. In this appendix, we focus on the no-interface case ($\lambda = 1$), where the factor $\alpha_k^2$ is absent. The question we would like to ask is whether the cubic scaling $g \sim 1/N^3$ is general for an arbitrary choice of the dissipative site $n_d$. In this case, the expression of the Liouvillian gap reduces to 
\be
g(\lambda=1) \approx \frac{\gamma}{2N+1} \min \left\{\sin^2\left( \frac{ n_d k\pi}{2N+1} \right) \right\} .
\ee
It is analytically hard to find the minimum for arbitrary $n_d$, however, by numerically running over all possible values of $k \in [1, 2N]$, for a given $N$ and $n_d \in [1, N]$, we find that the following minimum
\be
\min \left\{ q = n_d k \bmod (2N+1) \right\} 
\ee
has the solution
\be
q_\text{min} = 0 \text{ or } 1 .
\ee 

When the minium takes value of $q_\text{min} = 0$, we have $\sin^2 \left(\frac{n_d k \pi}{ 2N+1}\right) = 0$, indicating the presence of oscillating modes that never decay in the long-time lime. Especially, for certain choice of the system size and the dissipative site, there are extensive oscillating modes with vanishing real part in the corresponding Liouvillian spectrum. For example, when $N = 400$, $n_d = 267$, the number of oscillating modes is $266$, and the first non-decaying mode has $q = 267$. Such cases are not of interest for the purposes of the present work, but they would be useful in situations where we want to protect the initial information of the state.
In the discussion below, we will omit these cases for simplicity.

\begin{figure}[htp]
    \centering
    \begin{tikzpicture}
            \node[inner sep=0pt] (russell) at (-50pt,-85pt)
    {\includegraphics[width=.24\textwidth]{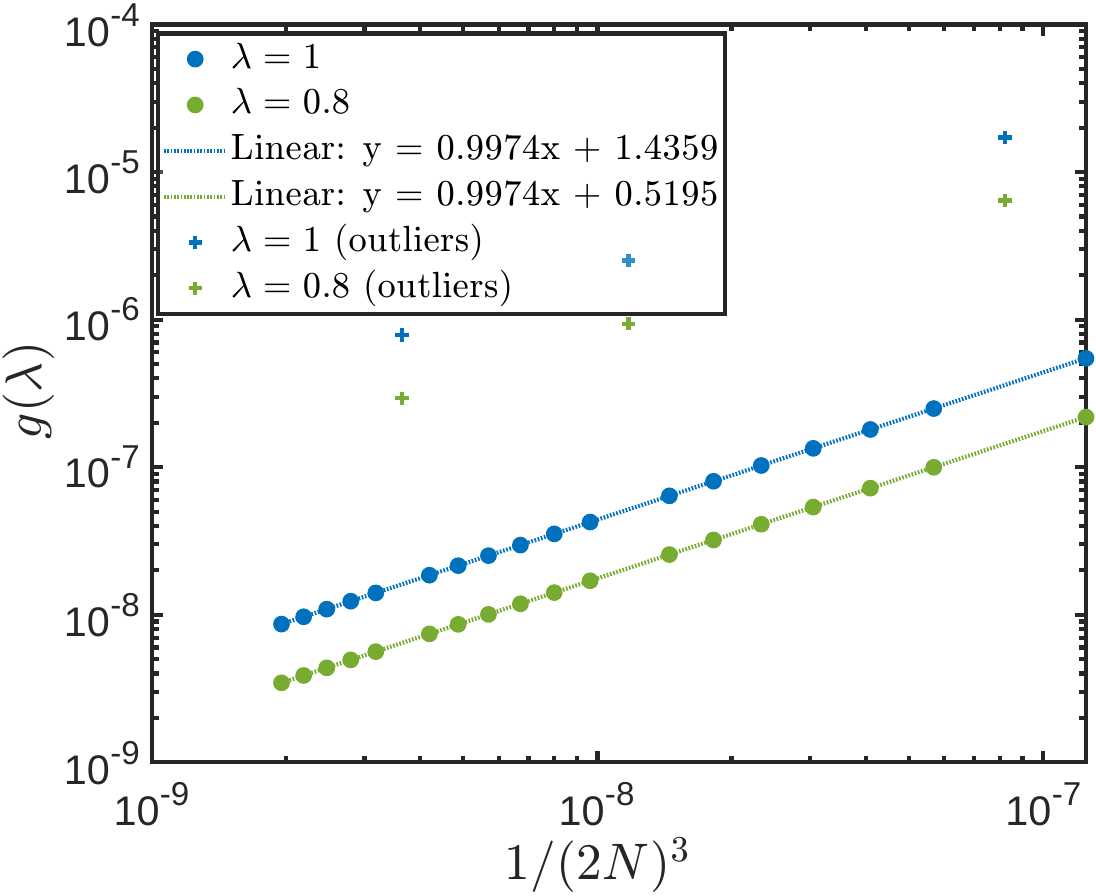}};
        \node[inner sep=0pt] (russell) at (75pt,-85pt)
    {\includegraphics[width=.245\textwidth]{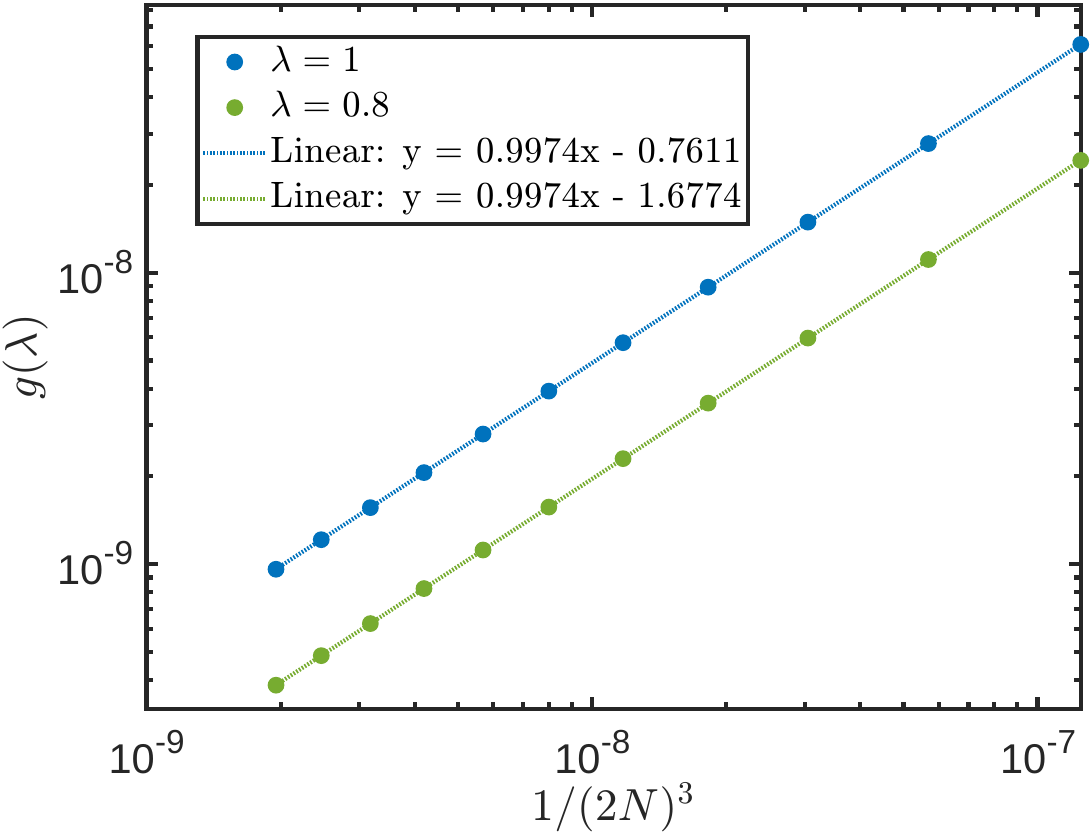}};

            \node at (-5pt, -105pt){(a)};
            \node at (120pt, -105pt){(b)}; 
\end{tikzpicture}
    \caption{The Liouvillian gap $g(\lambda)$ as a function of the total system size $L=2N$, with local dissipation applied at (a) $n_d = N-10$ and (b) $n_d = \frac{N}{2}$. For both the cases with ($\lambda = 0.8$) and without ($\lambda = 1$) conformal interface, we observe good cubic scaling, suggesting that it universally holds for general choices of the dissipative site $n_d$. The total system size is chosen as $2N = 200, 260, 320, \dots, 800$, and the dissipation strength is set to a small value of $\gamma = 0.05$.}
    \label{fig:general}
\end{figure}

Let us focus on the case where $q_\text{min} = 1$ (for example, the boundary dissipation $n_d = 1$ is such a case). In these cases, we have the Liouvillian gap in the absence of a conformal interface as
\be
g(\lambda=1) \approx \frac{\gamma}{2N+1} \sin^2 \left(\frac{ \pi}{2N+1}\right) . 
\ee
Under the limit $N \gg 1$, the above equation becomes
\be
g(\lambda=1) \approx \frac{\gamma \pi^2}{(2N+1)^3} \sim \frac{1}{N^3} ,
\ee
supporting the cubic scaling of the Liouvillian gap is universal for general choices of where the local dissipation is applied. (Note that there are still possibilities of occurring extensive oscillating modes for specific choices of the system size and dissipative site, where the sine function cannot be simply reduced to power-law scaling of $\sim 1/N$ by taking $N \to \infty$). In Fig.~\ref{fig:general}, we present numerical results of the Liouvillian gap for two choices of the dissipative sites $n_d = N-10$ and $n_d = N/2$, as a complementary to the cases reported in the main text. Here we observe that the numerical results exhibit a cubic scaling, as expected by the perturbative analysis.

For the case with conformal interface, the situation is more complicated and one cannot simply comment on the parity of $k$ to take even or odd integer, resulting in ambiguity of determining the factor $\alpha_k^2$. Nevertheless, for a large system size $N \gg 1$, one can expect that the (real part of the) Liouvillian spectrum is dense, at least for the cases without oscillating coherence. In such cases, a finite value of the interface parameter $\lambda$ should show a significant difference when choosing $k$ to be even or odd, while the change in the sine function is comparably small. Based on this, we argue that, to obtain the minimum of the Liouvillian gap, $k$ would take an odd integer, such that
\be
g(\lambda) \approx \frac{\gamma (1 - \sqrt{1 - \lambda^2})}{2N+1} \min \left\{ \sin^2 \left(\frac{ n_d k\pi}{2N+1}\right) \right\} . 
\ee
This suggests that the relation
\be
\frac{g(\lambda)}{g(1)} \approx 1 - \sqrt{1 - \lambda^2} 
\ee
holds for a general choice of the dissipative sites $n_d$, which is consistent with our numerical results shown in Fig.~\ref{fig:MainResult}. Moreover, we check the scaling behavior of the Liouvillian gap with the system size in the presence of conformal interface. As shown in Fig.~\ref{fig:general}, it displays a good cubic scaling $g \sim 1/N^3$.

\section{Dynamics of particle density}\label{app:dynamics}

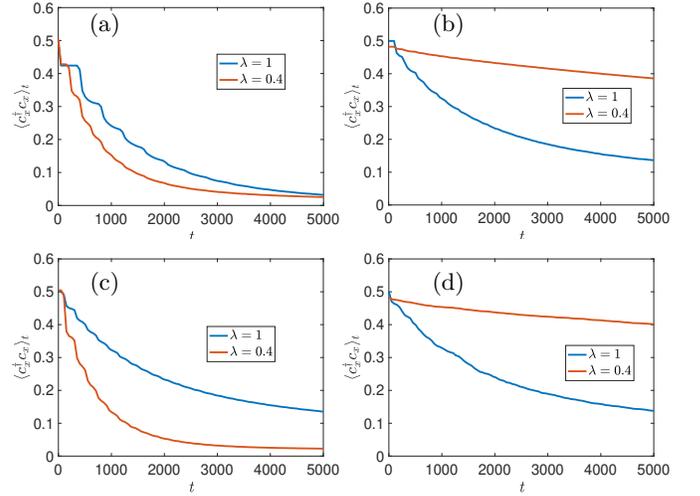
\begin{figure}[htp]
    \centering
    \begin{tikzpicture}
            \node[inner sep=0pt] (russell) at (-50pt, -85pt)
    {\includegraphics[width=.255\textwidth]{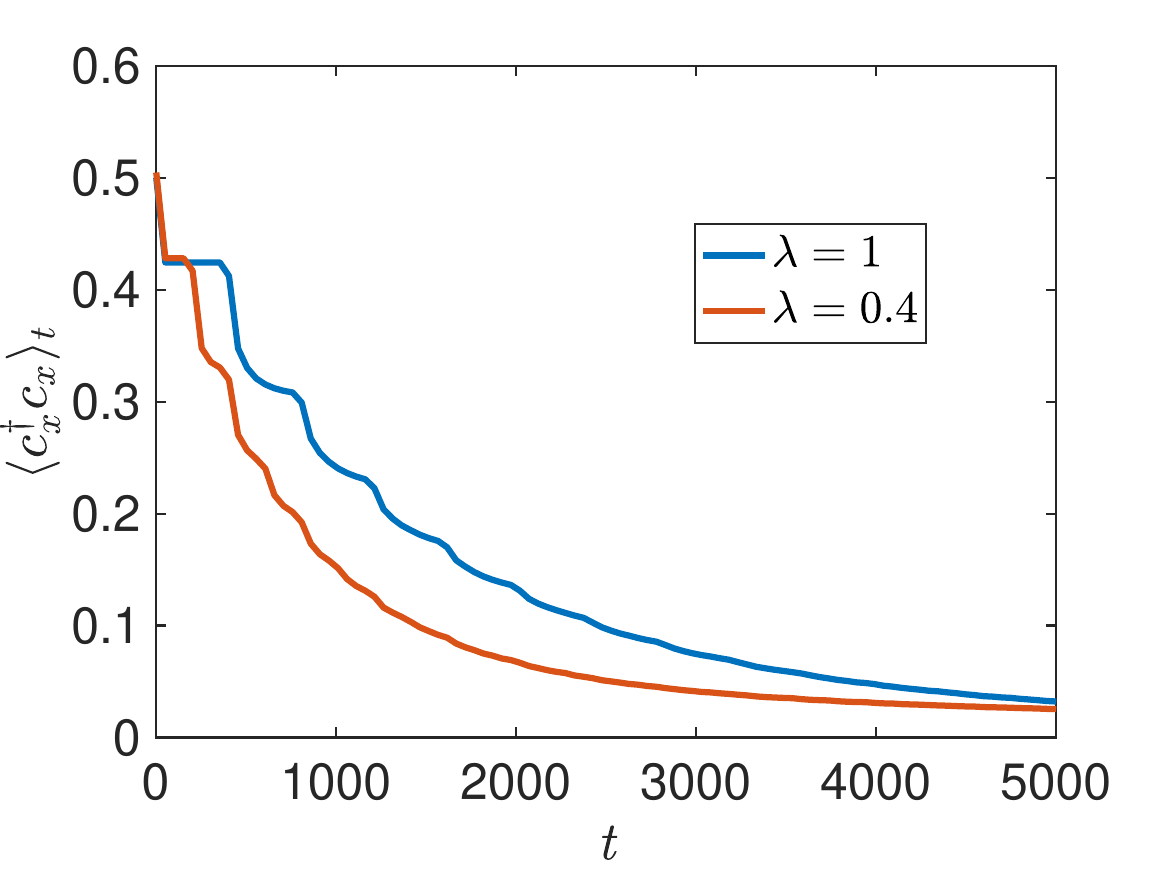}};
        \node[inner sep=0pt] (russell) at (75pt,-85pt)
    {\includegraphics[width=.255\textwidth]{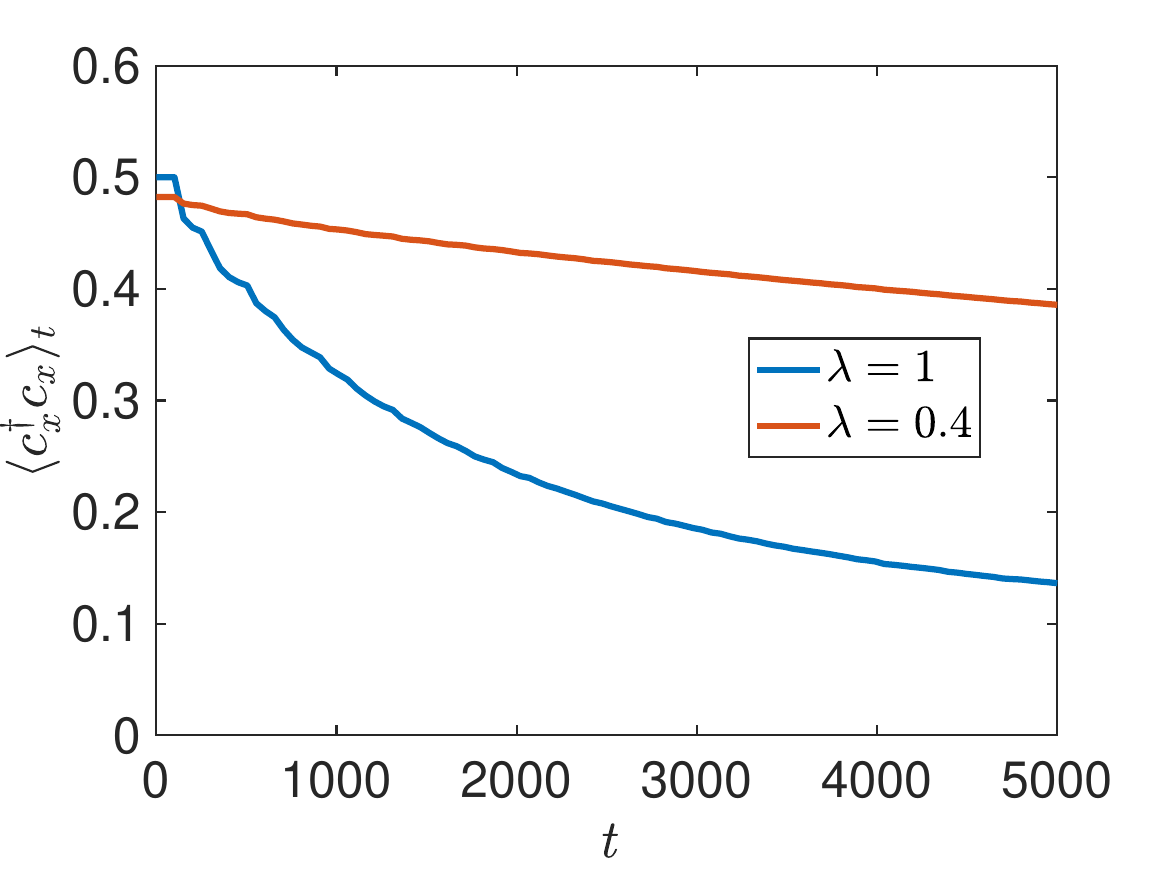}};
        \node[inner sep=0pt] (russell) at (-50pt, -180pt)
    {\includegraphics[width=.255\textwidth]{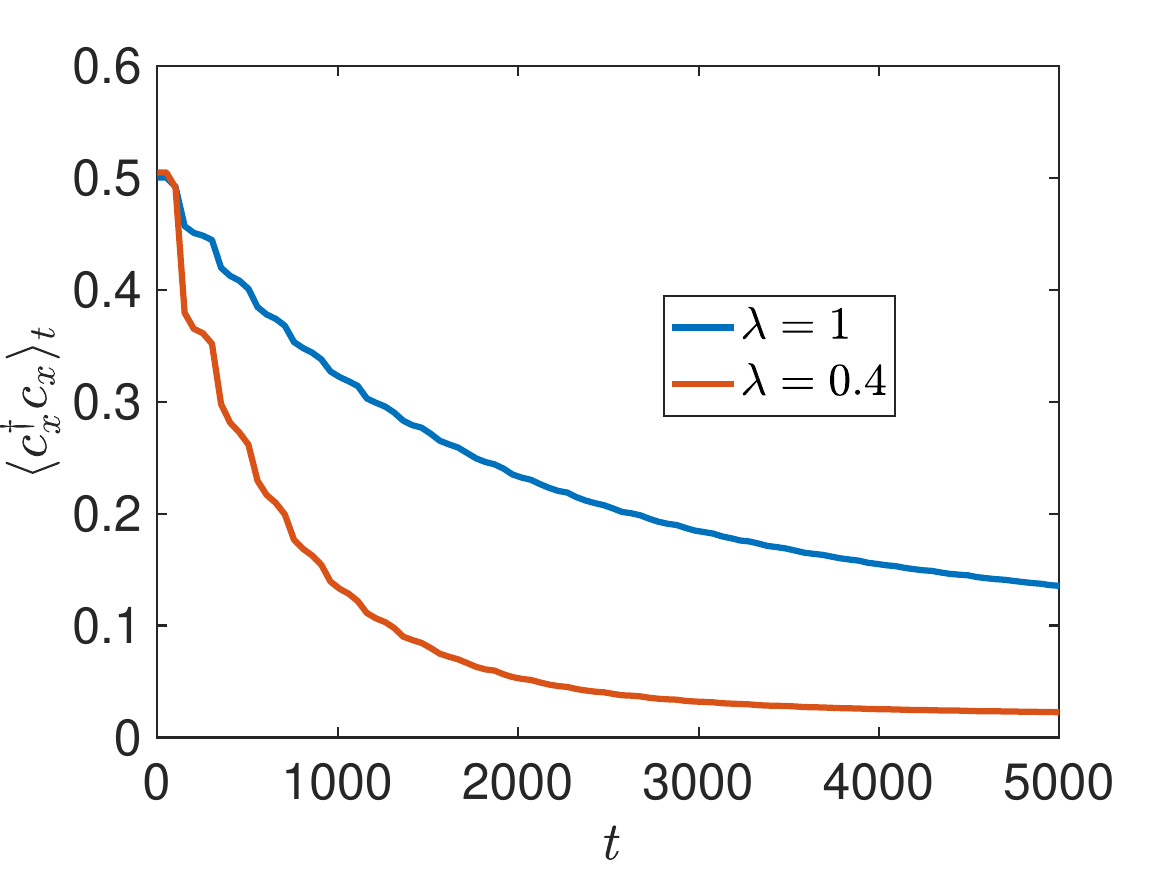}};
        \node[inner sep=0pt] (russell) at (75pt,-180pt)
    {\includegraphics[width=.255\textwidth]{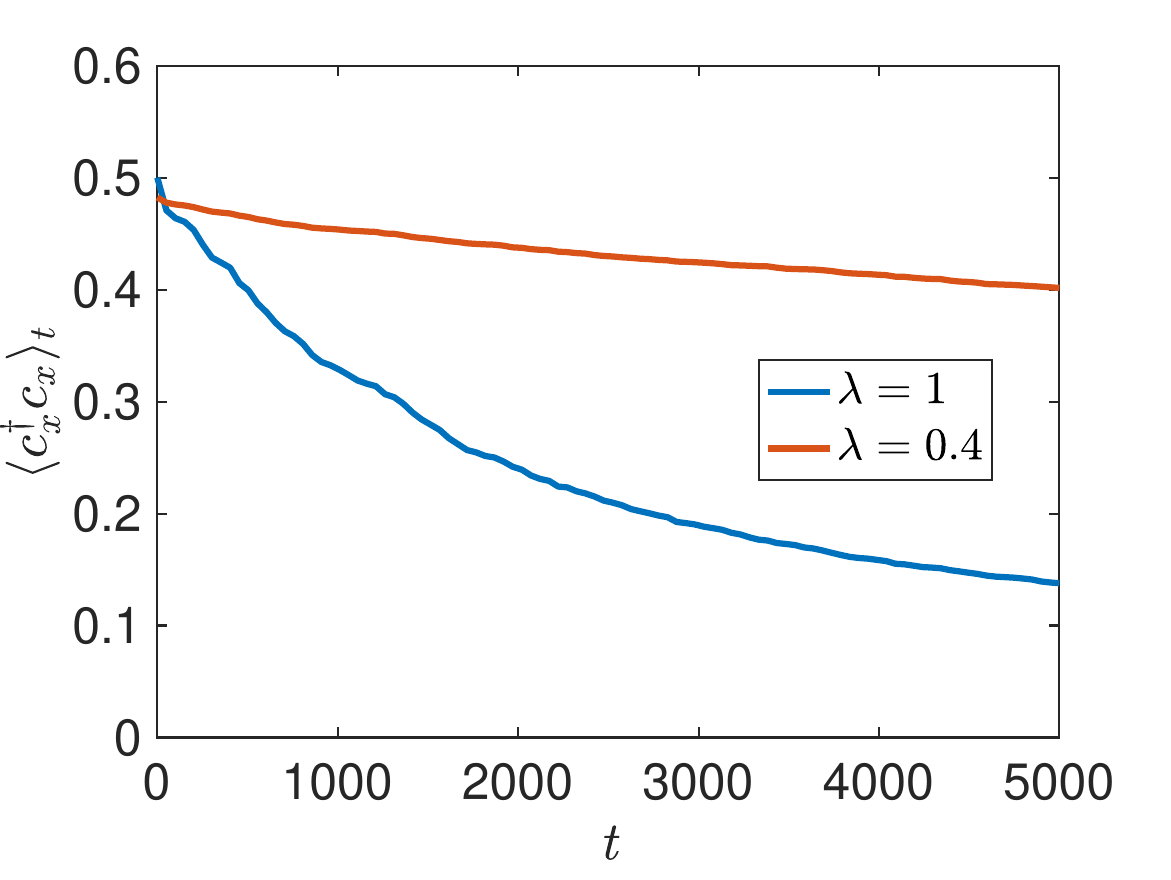}};
    
            \node at (-80pt, -50pt){(a)};
            \node at (50pt,-50pt){(b)};
            \node at (-80pt,-148pt){(c) }; 
            \node at (50pt,-148pt){(d)}; 
\end{tikzpicture}
    \caption{Time evolution of particle density $\langle c_x^\dag c_x\rangle_t$ in the presence of local dissipation at a site $n_d$ in the left half chain. (a) The dissipation is applied at the left boundary $n_d = 1$, and we examine the particle density at the same site $x = 1$. (b) The dissipation is applied at at the left boundary $n_d = 1$, and we examine the particle density at site $x = N+10 = 110$. (c) The dissipation is applied at the closest site to the interface $n_d = N-1$, and we examine the particle density at at the left boundary $x = 1$.  (d) The dissipation is applied at the closest site to the interface $n_d = N-1$, and we examine the particle density at site $x = N+10 = 110$. 
    Here, the total system is of length $2N = 200$ and the dissipation strength is set to $\gamma_{-} = 0.1$. Here the initial states are set to be the ground state of the interface Hamiltonian, therefore the initial values of the particle density varies with the interface parameter $\lambda$.}
    \label{fig: correl}
\end{figure}

\begin{figure}[htp]
   \centering
    \begin{tikzpicture}
            \node[inner sep=0pt] (russell) at (-50pt, -85pt)
    {\includegraphics[width=.255\textwidth]{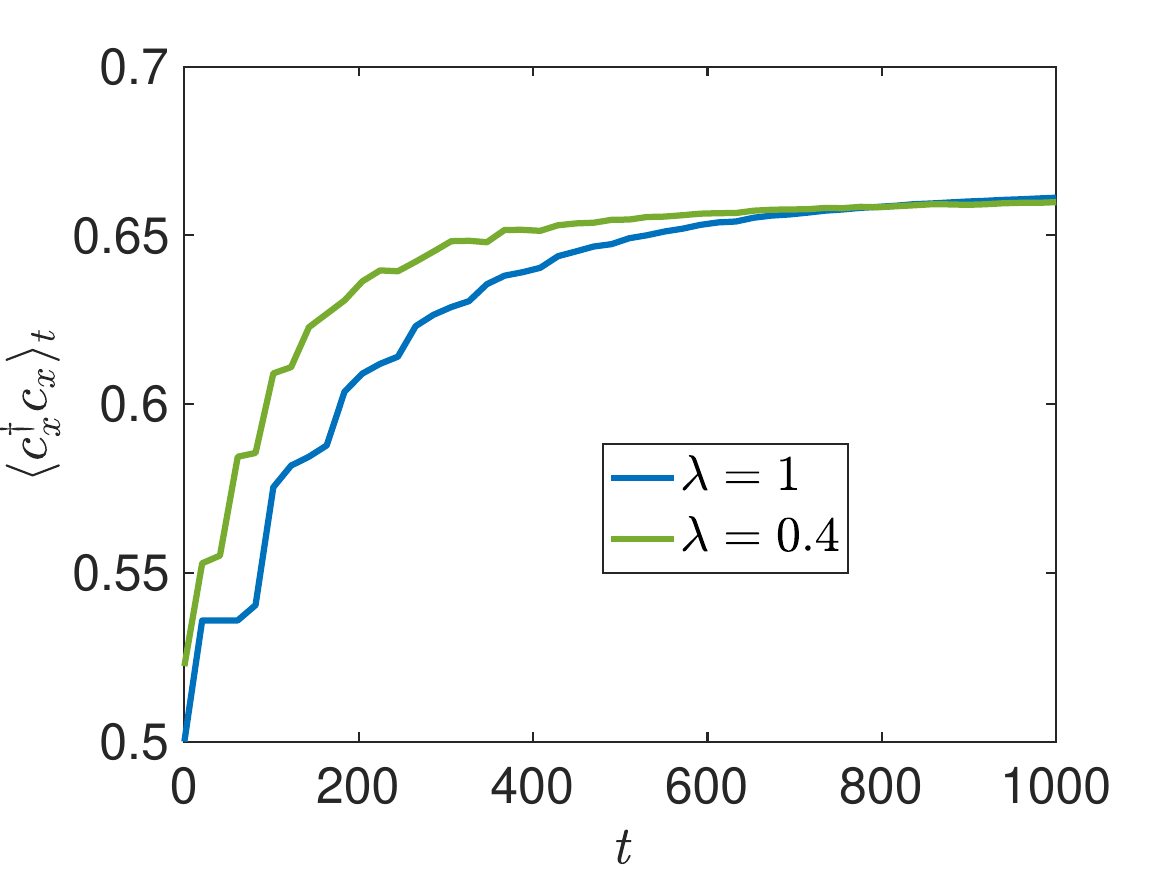}};
        \node[inner sep=0pt] (russell) at (75pt,-85pt)
    {\includegraphics[width=.255\textwidth]{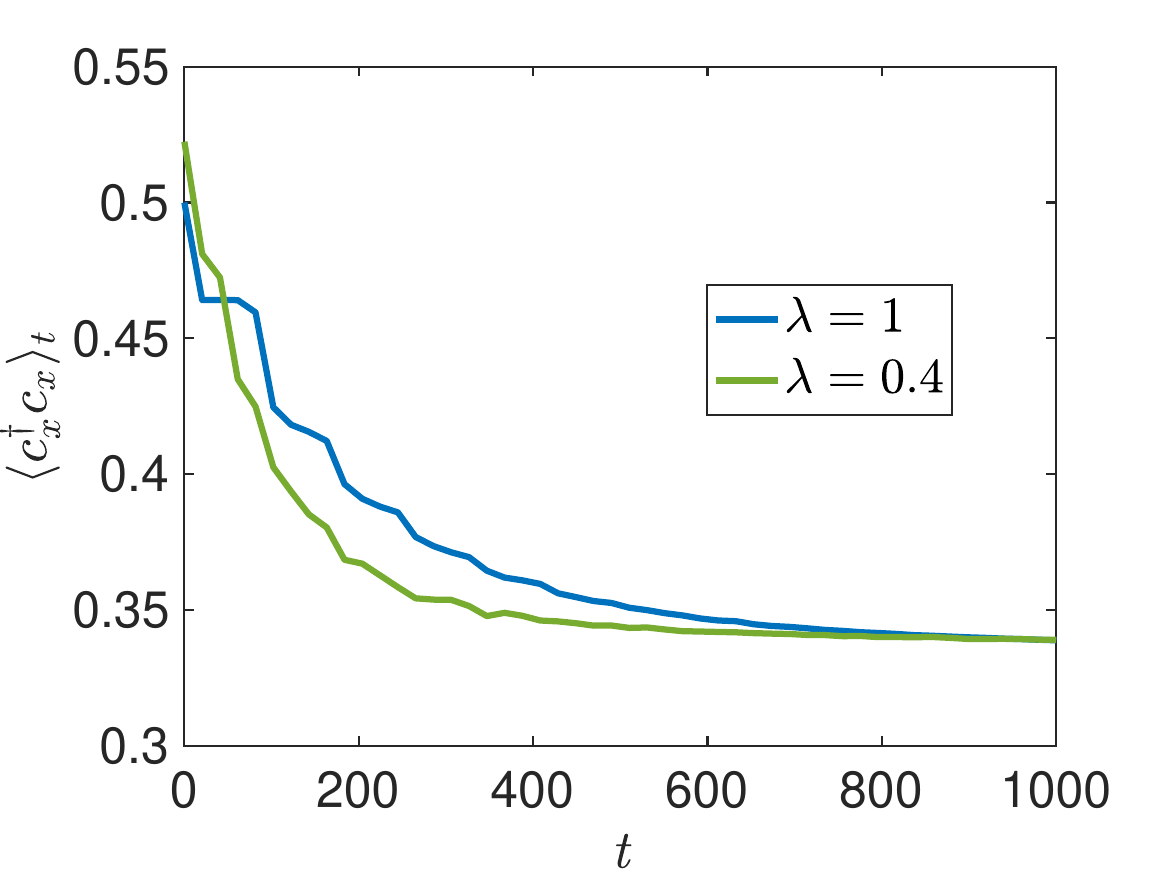}};
            \node at (-80pt, -50pt){(a)};
            \node at (55pt,-50pt){(b)};
\end{tikzpicture}
    \caption{Time evolution of particle density $\langle c_x^\dag c_x\rangle_t$ in the presence of local fermionic particle loss and gain both at the left boundary $n_d = 1$. Here we consider the particle density $n_x(t)$ for the left boundary $x=1$, which is the same location of the local dissipation with strength (a) $\gamma_{+}$ (0.1) > $\gamma_{-}$ (0.05) and (b) $\gamma_{+}$ (0.05) < $\gamma_{-}$ (0.1). The total system is of length $40$.} 
    \label{fig: gain and loss}
\end{figure}

In this appendix, we study the dynamics of particle density 
\be
\langle c_x^\dag c_x\rangle_t := \Tr [\rho(t) c_x^\dagger c_x]
\ee
at site $x$ under the time evolution driven by the Lindbladian superoperator in Eq.~\eqref{eq:lindblad} in the main text. The initial condition is chosen to be $\rho(t=0) = | \text{GS} \rangle \langle \text{GS} |$, where $| \text{GS} \rangle$ is the ground state of $H$ with the conformal interface as described in Eqs.~\eqref{eq:H1} and ~\eqref{eq:H2}. A particle loss and/or gain is then introduced at site $n_d$.  Here we include both losses and gains, with jump rates (dissipation strengths) $\gamma_-$ and $\gamma_+$, respectively [See \eqref{eq:Gamma}]. Since the Lindbladian superoperator is quadratic, the dynamics is exactly solvable, and the state remains Gaussian during the time evolution. In particular, the dynamics is fully described by the time-dependent correlation functions 
\be 
G_{x, y}(t) = \text{Tr}\left[ \rho(t)c^{\dagger}_{x}c_{y}\right] ,
\ee
where $x, y \in [1, 2N]$ are site indices, and $G$ is a $2N \times 2N$ matrix known as the correlation matrix. 
It can be calculated by~\cite{Fleischhauer2012, Alba2020, Tarantelli2021, Alba2023}
\be  \label{eq_app:Gt}
G(t) = e^{\textbf{H}_{\text{eff}} t} G(0) e^{\textbf{H}_{\text{eff}}^{\dagger} t}  +  \int^{t}_{0}dz e^{\textbf{H}_{\text{eff}} (t-z)}\Gamma^{+}e^{\textbf{H}_{\text{eff}}^\dagger (t-z)}. 
\ee 
where $\textbf{H}_{\text{eff}} = i\textbf{H} - \frac{1}{2} ( \mathbf{\Gamma}^+ + \mathbf{\Gamma}^- )$ is the (kernel of the) effective non-hermitian Hamiltonian as defined in Eq.~\eqref{eq:Heff_nonherm} in the main text (but including both gain and loss here), and $\mathbf{\Gamma}^\pm_{x,y} = \gamma_\pm \delta_{x,y} \delta_{x,n_d}$ describes the particle gain and loss at the same site $n_d$.

Based on this, we numerically calculate the dynamics of the particle density  $\langle c_x^{\dagger} c_x\rangle_t$, which is the diagonal element of the time-dependent correlation matrix $G(t)$.  We start with the case with only particle loss, i.e., $\gamma^- = \gamma, \gamma^+ = 0$, and the results are presented in Fig.~\ref{fig: correl}. In this case, the  In our setup, the conformal interface is located in the middle of the chain, and the local dissipation is applied on the left half-chain. Therefore, we expect the particle density exhibits dramatically different dynamical behavior on the two sides of the interface. As shown in Fig.~\ref{fig: correl} (a) and (c), when we look at the particle density in the left half-chain (where local dissipation is applied), the particle density decays more rapidly in the presence of an interface ($\lambda = 0.4$). In contrast, as shown in Fig.~\ref{fig: correl} (b) and (d), the particle density in the right half-chain decays more rapidly in the absence of an interface ($\lambda = 1$). We have also checked that the particle density at a position close to the interface $x = N+10$ and at the right boundary $x = 2N$ exhibits similar relaxation dynamics. Intuitively, this can be understood by the partial reflection and transmission of the quasi-particles across the interface~\cite{Alba2022}, which leads to a shorter effective length of the left half-chain but a suppression of the relaxation rate of the right half-chain. 
Moreover, in Figure~\ref{fig: gain and loss}, we also investigate the dynamics of particle density $\langle c_x^{\dagger} c_x\rangle_t$ in the case where both local particle loss and gain are applied at the left boundary of the system ($n_d = 1$), with the dissipation strength $\gamma_-$ and $\gamma_+$ respectively. Specifically, we consider the particle density at the left boundary $x = 1$, which is the same position where the dissipation is applied. As shown in Fig.~\ref{fig: gain and loss}, for both cases of $\gamma_+ > \gamma_-$ (see panel (a)) and $\gamma_+ < \gamma_-$ (see panel (b)), the onsite particle number exhibits a sharper change in the presence of a conformal interface, similar to the behavior observed in the loss-only case shown in Fig.~\ref{fig: correl} (a) and (c).

\bibliography{defectCFT}

\end{document}